\documentclass[aps,prx,reprint,superscriptaddress,longbibliography,a4paper,raggedbottom,nofootinbib]{revtex4-2}

\usepackage[T1]{fontenc}
\usepackage{newtxtext}
\usepackage{amsmath}
\usepackage{amssymb}
\usepackage{amsthm}
\usepackage{capt-of}
\usepackage{newtxmath}
\usepackage{bm}
\usepackage{booktabs}
\usepackage{enumitem}
\usepackage{xcolor}
\usepackage{microtype}
\usepackage{graphicx}
\usepackage{placeins}

\usepackage{hyperref}
\hypersetup{colorlinks=true,linkcolor=blue,citecolor=blue,urlcolor=blue,filecolor=blue}
\usepackage{cleveref}

\makeatletter
\renewcommand{\section}{%
  \@startsection{section}{1}{\z@}%
    {10pt \@plus 2pt \@minus 1pt}{6pt}%
    {\normalfont\small\bfseries\centering}%
}
\renewcommand{\subsection}{%
  \@startsection{subsection}{2}{\z@}%
    {10pt \@plus 2pt \@minus 1pt}{4pt}%
    {\normalfont\small\bfseries\centering}%
}

\AtBeginDocument{%
  \g@addto@macro\bibfont{\interlinepenalty=10000\relax}%
}
\makeatother

\newcommand{\SO}[1]{\mathrm{SO}(#1)}
\newcommand{\SU}[1]{\mathrm{SU}(#1)}
\newcommand{\U}[1]{\mathrm{U}(#1)}
\newcommand{\half}{\tfrac{1}{2}}
\newcommand{\lsh}{\ell_{\mathrm{sh}}}
\newcommand{\Nbr}[1]{\mathcal{N}(#1)}

\newcommand{\best}[1]{\textbf{#1}}
\newcommand{\secondbest}[1]{\underline{#1}}
\newcommand{\apxref}[1]{\hyperref[#1]{Appendix~\ref*{#1}}}

\makeatletter
\AtBeginDocument{%
  \let\arxiv@hangfrom@section\@hangfrom@section
  \let\arxiv@sectioncntformat\@sectioncntformat
}
\makeatother

\begin{document}

\title{SU(2)-Equivariant Spinor Message Passing for Molecular Magnetism}

\author{Erh-Wei Sheng}
\email[Contact author: ]{ewinnie@gapp.nthu.edu.tw}
\affiliation{Division of Engineering and Applied Science, California Institute of Technology,
Pasadena, California 91125, USA}
\affiliation{Department of Chemical Engineering, National Tsing Hua University,
Hsinchu 300044, Taiwan}

\author{Beom Seok Kang}
\affiliation{Division of Chemistry and Chemical Engineering, California Institute of Technology,
Pasadena, California 91125, USA}

\author{Yuchao Lin}
\affiliation{Division of Engineering and Applied Science, California Institute of Technology,
Pasadena, California 91125, USA}

\author{Danish Khan}
\affiliation{Division of Engineering and Applied Science, California Institute of Technology,
Pasadena, California 91125, USA}

\author{Anima Anandkumar}
\email[Contact author: ]{anima@caltech.edu}
\affiliation{Division of Engineering and Applied Science, California Institute of Technology,
Pasadena, California 91125, USA}

\begin{abstract}
Predicting molecular magnetic properties often requires costly
electronic-structure calculations that account for spin--orbit coupling.
Inspired by the spinor nature of the underlying electronic states, we
introduce eSpinor, an $\SU{2}$-equivariant graph neural network that
propagates spinor features alongside the integer-rank tensors used in
conventional equivariant models. We initialize spinors from eigenvectors
of an approximate Hamiltonian that includes spin--orbit coupling.
The network preserves their rotation laws and makes predictions independent
of arbitrary eigenbasis choices. Comparisons with geometry-only and
Hamiltonian-informed baselines assess the benefit of this architecture
beyond supplying electronic information. On cobalt single-molecule magnet
candidates, eSpinor improves magnetic anisotropy predictions, including for
unseen ligands. It also improves excitation-energy predictions for dysprosium
candidates with unseen ligands, extending the benefit to rotationally invariant
scalar targets. For a dysprosium magnetic resonance imaging probe, it
predicts real electron-spin correlations and low-field nuclear relaxation
rates more accurately despite static crystal-field errors similar to the
Hamiltonian-informed baseline. This dynamical advantage extends from water
to methanol without retraining. These results support spinor message
passing as a useful inductive bias for molecular magnetism in the tested
systems and show why model selection should be guided by accuracy on the
intended physical observables.
\end{abstract}

\maketitle

\section{Introduction}
\label{sec:intro}

Molecular magnetism underlies single-molecule magnets
(SMMs)~\cite{craig2015sim3d,goodwin2017dysprosocenium}, molecular qubits for
quantum information~\cite{gaita2019qubits}, and magnetic resonance imaging
(MRI) probes~\cite{harvey2013parashift}. Designing these systems requires
predicting how ligand chemistry and geometry affect magnetic energy levels
and dynamics. Relevant quantities include zero-field-splitting (ZFS)
parameters in transition-metal complexes and crystal-field parameters in
lanthanide ions~\cite{boca2004zfs,stevens1952operators}.
The ligand field and spin--orbit coupling (SOC), the relativistic
interaction between an electron's orbital and spin angular momenta,
shape the magnetic anisotropy of these
complexes~\cite{abragam1970epr}. Changes in molecular geometry can
therefore influence both their magnetic energy levels and spin
dynamics~\cite{thompson2026parashift}.

A common approach uses complete active space self-consistent field
spin--orbit (CASSCF-SO) calculations~\cite{thompson2026parashift}.
CASSCF first describes the multireference electronic structure without
SOC~\cite{roos1980casscf}. SOC is then included by mixing the resulting
electronic states in a state-interaction
calculation~\cite{malmqvist2002rassi}. Unlike density functional
theory (DFT), which is single-reference and scales polynomially with system
size, CASSCF captures the near-degenerate multiplets of open-shell metal
ions by including all electronic configurations within an active space.
The number of these configurations grows combinatorially with the size of
the active space, which restricts conventional calculations to small active
spaces~\cite{levine2020casscf}. Averaging over many low-lying states adds
further cost, so a single geometry can require hours of computation
(Sec.~\ref{sec:discussion-observables}), making repeated calculations for
screening or dynamics expensive. Machine learning surrogates that predict
magnetic parameters from geometry offer a cheaper alternative. For example,
Gaussian-moment neural networks (GM-NN) have been extended to predict
magnetic anisotropy tensors~\cite{zaverkin2022gmnn}.

Other models incorporate spin explicitly. For magnetic materials, SpinGNN
uses atomic positions and magnetic-moment directions to learn spin--lattice
potential-energy surfaces~\cite{yu2024spingnn}. SpinGNN++ extends this
approach to SOC-dependent interactions while enforcing spatial and
time-reversal symmetries~\cite{yu2024time}. For open-shell molecules,
AIMNet2-NSE predicts energies using neural spin-charge equilibration
constrained by total charge and spin multiplicity~\cite{kalita2026aimnet2nse}.
OrbitAll learns from spin-polarized orbital matrices computed with
semiempirical methods~\cite{kang2025orbitall}. Other approaches predict
electronic Hamiltonian matrices, including SOC, directly from atomic
structure, as in DeepH-E3, Uni-HamGNN, and
MACE-H~\cite{gong2023deephe3,zhong2026unihamgnn,qian2026maceh}.

Several of these models impose equivariance under $\SO{3}$, the group of
proper rotations in three-dimensional space, so that predicted tensors
rotate consistently with the molecular geometry~\cite{duval2023hitchhiker}.
This inductive bias can improve data efficiency and accuracy. In
machine-learning interatomic potentials (MLIPs), nonscalar
$\SO{3}$-equivariant features help even for rotation-invariant
energies~\cite{batzner2022nequip}, showing that internal features can
usefully carry richer rotation laws than the outputs.
For magnetic-property prediction, eqSNAP provides an $\SO{3}$-equivariant
extension of the spectral neighbor analysis potential (SNAP) method.
It has been used to predict ZFS tensors and crystal-field parameters using
local atomic descriptors and linear
regression~\cite{nguyen2022eqsnap,briganti2025spinphonon}.

\begin{figure*}[t]
  \centering
  \includegraphics[width=\textwidth]{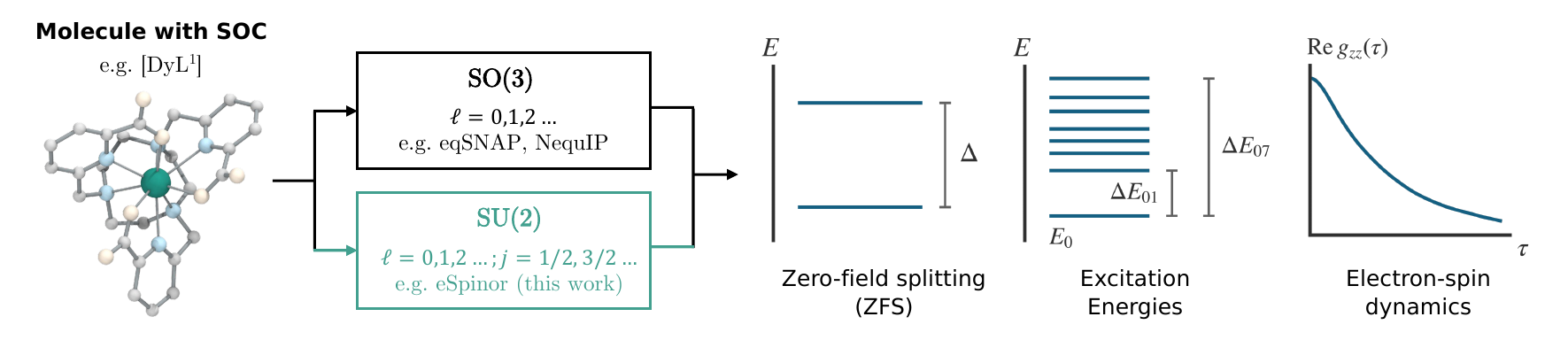}
  \caption{Representation choices for predicting SOC-dependent molecular magnetic properties.
  eSpinor propagates half-integer spinor features alongside integer-rank
  tensor features. The ZFS levels are schematic, and the electron-spin correlation
  $\operatorname{Re}g_{zz}(\tau)$ shows the $3$~T CASSCF-SO reference~\cite{thompson2026parashift}.}
  \label{fig:overview}
\end{figure*}

SOC couples orbital motion and electron spin, motivating representations
that describe these degrees of freedom jointly~\cite{abragam1970epr}.
Although magnetic observables follow tensor rotation laws, the underlying
electronic states can transform as spinors under
$\SU{2}$~\cite{littlejohn2021rotations,littlejohn2021coupling}.
By analogy with MLIPs, such features could serve as internal features
and benefit the predictions of integer-rank magnetic targets, 
yet the models discussed above use only integer-rank internal features.
This motivates our research question, illustrated in Fig.~\ref{fig:overview}:
\textit{Can $\SU{2}$-equivariant spinor message passing provide a useful
inductive bias for predicting SOC-dependent molecular properties?}

Inspired by the spinor nature of spin--orbit-coupled electronic states, we
develop eSpinor, an $\SU{2}$-equivariant graph neural
network that propagates spinor features alongside integer-rank
tensors. We initialize spinors from eigenvectors of an approximate SOC
Hamiltonian and classify the allowed interaction types to guarantee
rotation equivariance and eigenbasis independence. 
To assess the proposed architecture, we compare eSpinor with 
the widely used $\SO{3}$-equivariant architecture NequIP~\cite{batzner2022nequip} 
and its Hamiltonian-informed variant NequIP+H, which serves as an ablation. 
On cobalt SMM
candidates~\cite{mariano2024compass}, eSpinor generalizes better to unseen
ligands. The improvement also extends to rotationally invariant scalar
excitation energies in Dy(III) complexes with unseen
ligands~\cite{frangoulis2026dysmm}. On a dysprosium MRI probe~\cite{thompson2026parashift}, it predicts
real electron-spin correlations more accurately at 8k training geometries,
even though it matches NequIP+H on static crystal-field tensors at individual
geometries. The dynamical advantage also extends from water to methanol
without retraining. These results indicate that a model with
$\SU{2}$-equivariant spinor features improves magnetic-property prediction
in the tested systems for both equivariant tensors and invariant scalars.
They also show that similar static tensor
errors can lead to different dynamical errors. Model selection should
therefore be guided by accuracy on the physical observables relevant to
the intended application.

\begin{figure*}[t]
  \centering
  \includegraphics[width=\textwidth]{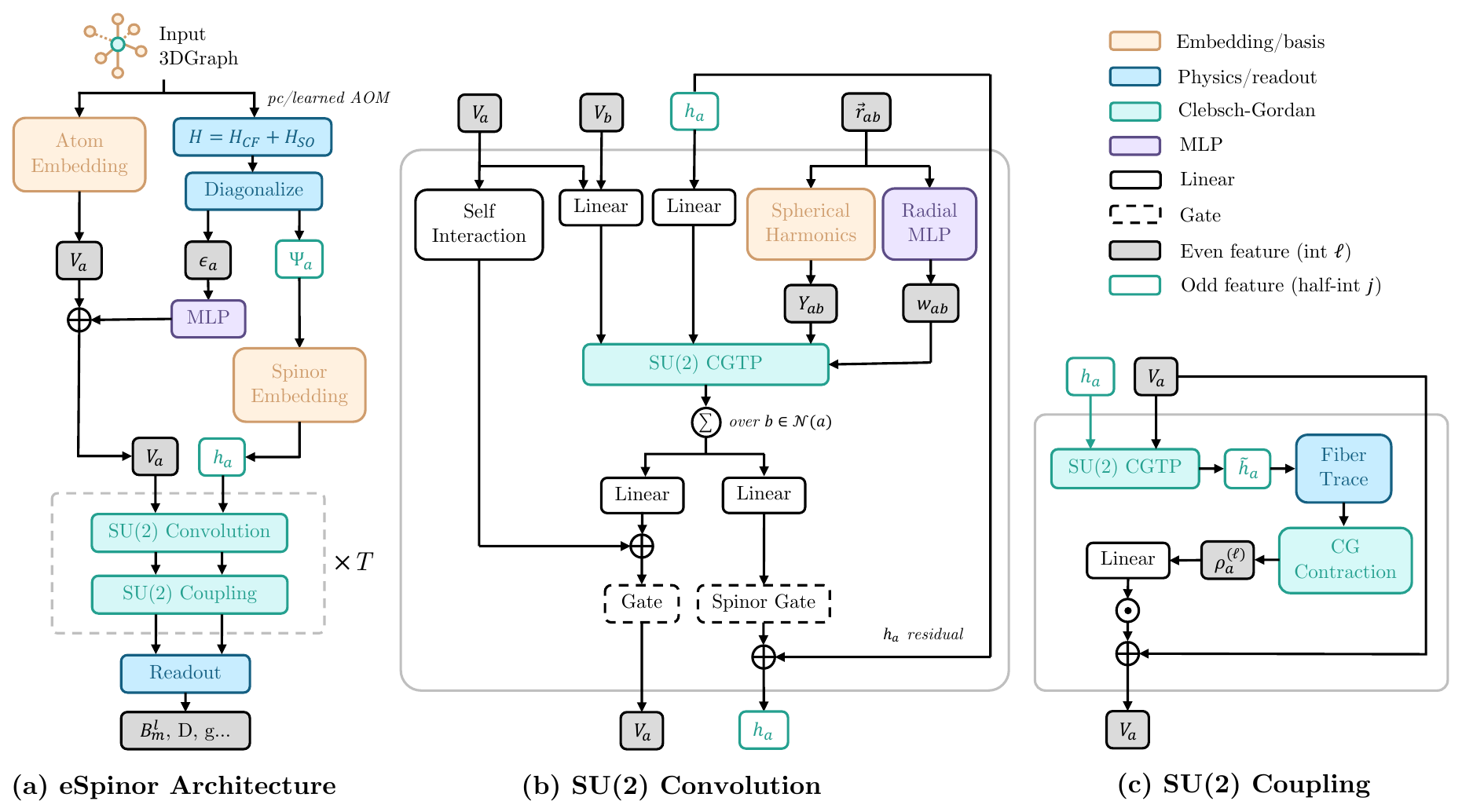}
  \caption{eSpinor architecture. (a) Atomic embeddings and Hamiltonian
  eigenvectors initialize integer tensors $V_a$ and spinors $h_a$ for
  repeated message passing and tensor readout.
  (b) Convolution uses Clebsch--Gordan tensor products (CGTP).
  (c) Spinor contractions update integer tensors.
  Here even/odd denotes integer/half-integer rank and $T=N_{\mathrm{lay}}$.
  Tensor benchmarks use $\mathbf D$ and $B_m^\ell=B_q^{(\ell)}$.
  Excitation energies use an invariant readout described in
  Sec.~\ref{sec:dy-spectra}.
  The preliminary tensor convolution used by the learned AOM initializer
  is omitted.}
  \label{fig:espinor-architecture}
\end{figure*}

\section{SU(2)-Equivariant Spinor Message Passing}
\label{sec:spinor-method}

Our first task is to construct spinor features that transform
consistently when the molecular geometry rotates
(Sec.~\ref{sec:initialization}). Their initialization introduces a freedom
in choosing the eigenbasis, which constrains the network design
(Sec.~\ref{sec:gauge}). We then design message-passing layers that
respect these constraints (Sec.~\ref{sec:couplings}) and a readout that
produces magnetic tensors independent of the eigenbasis choice
(Sec.~\ref{sec:tensor-readout}).

Figure~\ref{fig:espinor-architecture}(a) summarizes this construction.
We represent the molecule as a graph $G$ with atomic numbers $Z_a$ and
positions $\mathbf r_a$. Edges connect atoms within a radial cutoff.
Every atom carries integer-rank tensor features $V_a$, initialized from
atomic-species embeddings. We consider SOC only at the magnetic metal sites, where we initialize 
spinor features $h_a$ from
an approximate Hamiltonian that combines SOC with the ligand field
(Sec.~\ref{sec:initialization}).
Thus $V_a$ and $h_a$ are two feature types attached to the same graph.
Within each interaction layer, the
convolution in Fig.~\ref{fig:espinor-architecture}(b) updates both feature types.
The coupling in Fig.~\ref{fig:espinor-architecture}(c) then contracts spinors
to supply additional integer tensors while retaining the spinors for the
next layer. A final readout produces magnetic tensors independent of the
basis chosen for the initial eigenvectors.

\subsection{Spinor initialization}
\label{sec:initialization}

To specify the two feature types, we use irreducible representations
(irreps). 
We label integer-rank tensor features by $\ell=0,1,\ldots$ and 
spinor features by $j=\half,\tfrac32,\ldots$ which transform under half-integer irreps of $\SU{2}$, the double cover of $\SO{3}$~\cite{littlejohn2021rotations,zee2016group}.
  
Integer-rank geometric features cannot generate spinors through tensor
products~\cite{varshalovich1988quantum}. We therefore introduce electron
spin explicitly by using the single-electron spin--orbital space of a metal
shell. The orbital and spin angular momenta $\mathbf L$ and $\mathbf S$
co-rotate under joint spatial rotations generated by
$\mathbf J=\mathbf L+\mathbf S$. Combining the shell's integer orbital
angular momentum $\lsh$ with electron spin $s=\half$ gives the half-integer
irreps $j=\lsh\pm\half$~\cite{littlejohn2021coupling}.
To obtain spinor features that depend on the ligand environment and SOC,
we construct an approximate Hamiltonian at each metal site $a$,
\begin{equation}
H_{\mathrm{init},a}(G)
=H_{\mathrm{CF},a}(G)\otimes I_2+\xi_a\mathbf L\cdot\mathbf S.
\label{eq:initial-hamiltonian}
\end{equation}
Here $H_{\mathrm{CF},a}$ acts on orbital states and $I_2$ on spin.
The SOC strength is $\xi_a$, with $\mathbf L$ and $\mathbf S$ in units of
$\hbar$~\cite{abragam1970epr,lueken2008lectures}.
This zero-field, single-electron Hamiltonian initializes features for
targets obtained from many-electron calculations.

Specifically, we choose $H_{\mathrm{CF},a}$ to reflect the metal--ligand
bonding in each benchmark.
For Dy(III), the relatively shielded $4f$ shell motivates a point-charge
model based on ligand positions and effective
charges~\cite{abragam1970epr,waldmann2025chargepotential}.
For Co(II), we compare point-charge initialization with a learned
angular-overlap model (AOM) to account for metal--ligand
covalency~\cite{schaffer1965aom}.
The initializer formulas and numerical parameters are given in the
Supplemental Material~\cite{supplementalMaterial}
(Sec.~\ref{supp-sec:initializer-details}).

Diagonalizing the Hamiltonian gives approximate electronic energy levels
$\epsilon_{a,u}$ and their corresponding eigenvectors. We collect the
$k=2(2\lsh+1)$ orthonormal eigenvectors as columns of $\Psi_a$ and their
energy levels in $E_a=\operatorname{diag}(\epsilon_{a,1},\ldots,\epsilon_{a,k})$,
so the eigenproblem is
\begin{equation}
H_{\mathrm{init},a}\Psi_a=\Psi_a E_a.
\label{eq:initial-eigensystem}
\end{equation}
We express these eigenvectors in the coupled basis $|j,m\rangle$, where
$m=-j,\ldots,j$ is the total angular momentum projection. At each $j$, their $2j+1$ components form spinors with Wigner
rotation matrices $D^{(j)}$, as derived in \apxref{app:spinor-symmetry}.
As shown in Fig.~\ref{fig:espinor-architecture}(a), the energy levels supply
invariant scalar features and the eigenvectors initialize spinor features.
Using these eigenvectors requires accounting for their eigenbasis freedom,
as explained in Sec.~\ref{sec:gauge}.

\subsection{Addressing eigenbasis gauge freedom}
\label{sec:gauge}

Diagonalization in Eq.~\eqref{eq:initial-eigensystem} determines eigenstates
up to unitary transformations within each eigenspace, giving an
eigenbasis gauge freedom. In our zero-field, single-electron
initializer, time-reversal symmetry produces degenerate Kramers
pairs~\cite{abragam1970epr,kramers1930rotation,wigner1932zeitumkehr}.
The eigenbasis of each energetically isolated pair $p$ is therefore defined
up to $Q_{a,p}\in\U{2}$, which includes individual $\U{1}$ phase choices
and unitary mixing of the 2 eigenvectors. Predictions must be independent
of this choice, a requirement also encountered in spectral graph
learning~\cite{lim2023signbasis,lim2023signequivariant}.
In \apxref{app:spinor-symmetry}, we give the conditions for larger
degenerate eigenspaces.

This gauge freedom also enters the transformation under molecular rotations.
For a rotation $g\in\SO{3}$ and either lift $\tilde g\in\SU{2}$, the
eigenvectors $\Psi_a$ obey
\begin{equation}
\begin{aligned}
\Psi_a(g\cdot G)&=U_{\mathrm{sh}}(\tilde g)\Psi_a(G)Q_a,\\
Q_a&=\operatorname{diag}(Q_{a,1},\ldots,Q_{a,k/2}).
\end{aligned}
\label{eq:initial-eigenbasis-covariance}
\end{equation}
Here the spin--orbital rotation $U_{\mathrm{sh}}$ acts through $D^{(j)}$
on the angular index $m$ at each $j$, while the gauge transformation $Q_a$ mixes eigenvector
columns $u$ within each pair. Following gauge-equivariant
learning~\cite{cohen2019gauge,huang2025gauge}, we preserve gauge covariance
in the spinor features alongside rotation equivariance. The spinor features $h_a$ remain basis dependent, but contracting their eigenbasis index $u$ gives the gauge-invariant product
$h_a h_a^\dagger$. We project this product onto integer-rank tensors.

For initialization, we use Boltzmann-form weights
$w_u\propto e^{-\bar\epsilon_u/T_B}$, where $\bar\epsilon_u$ is the
energy averaged within the Kramers pair containing eigenstate $u$.
The learned energy scale $T_B$ controls how strongly lower-energy
states are favored. This energy dependence allows contractions of
the eigenvectors to retain information about the ligand environment.
To preserve gauge covariance, the weights and invariant gates
act identically on both columns of each Kramers pair and therefore commute
with $Q_a$. Equivariant channel maps then give initial spinor features
at layer $n=0$ that satisfy
\begin{equation}
h_a^{(j),0}\mapsto D^{(j)}(\tilde g)h_a^{(j),0}Q_a.
\label{eq:spinor-initial-law}
\end{equation}
In \apxref{app:spinor-symmetry}, we derive this transformation law
from the initialization map. The weighting formulas and gate inputs are
given in the Supplemental Material~\cite{supplementalMaterial}.
Each subsequent layer $n$ must preserve Eq.~\eqref{eq:spinor-initial-law},
which guides the message passing in Sec.~\ref{sec:couplings}.

\subsection{Gauge-covariant message passing}
\label{sec:couplings}
\label{sec:architecture}

Each interaction layer updates the features using information from
neighboring atoms~\cite{gilmer2017mpnn}. We choose NequIP as the backbone
because it uses spherical harmonics to encode interatomic directions and
explicit Clebsch--Gordan tensor products, which can be directly extended to
half-integer representations~\cite{batzner2022nequip}.
The spinor updates must also preserve the eigenbasis gauge law in
Eq.~\eqref{eq:spinor-initial-law}.

\begin{samepage}
We classify the allowed real-bilinear couplings into 3 interaction types,
\begin{equation}
\label{eq:three-couplings}
V\otimes V\to V,\qquad
h\otimes V\to h,\qquad
h\otimes\overline h\xrightarrow{\,\sum_u\,}V.
\end{equation}
\end{samepage}
Here $\overline h$ denotes elementwise complex conjugation, and each product
includes Clebsch--Gordan projection onto allowed angular ranks.
The product $V\otimes V$ couples integer-rank features to produce
integer-rank tensors. The mixed product $h\otimes V$ updates spinor features
without changing the eigenbasis index $u$, so its output carries the same
$Q_a$. In the conjugated product $h\otimes\overline h$, contraction over
$u$ cancels the unitary $Q_a$ while retaining angular information in
gauge-invariant integer tensors.
We derive this classification and its covariance properties in
\apxref{app:spinor-symmetry}.

We use these operations to assemble the convolution in
Fig.~\ref{fig:espinor-architecture}(b). In the equations below, square
brackets $[\cdot]$ denote Clebsch--Gordan projection onto the output irreps, with angular-component and feature-channel indices omitted for clarity.
Let $Y_{ab}=Y(\hat{\mathbf r}_{ab})$ collect the spherical harmonics of the
direction from atom $a$ to atom $b$. The sum below runs over neighbors
$b\in\Nbr{a}$. In our mononuclear benchmarks,
only the metal carries spinor features, so ligand neighbors condition an
update of that metal's own $h_a$. We combine these neighbor contributions
with an on-site coupling to $V_a^n$ in a gated residual update,
\begin{subequations}
\label{eq:interaction-layer}
\begin{equation}
h_a^{n+1}=h_a^n+\sigma_h^n\Bigl(
\sum_b A_{ab}^n[h_a^n\otimes Y_{ab}]
+W_{\mathrm{on}}^n[h_a^n\otimes V_a^n]\Bigr).
\label{eq:spinor-residual-update}
\end{equation}
Here the equivariant channel maps are absorbed into the neighbor weights
$A_{ab}^n$ and the on-site coupling $W_{\mathrm{on}}^n$.
The neighbor weights depend on distance and invariant scalar features at
both atoms and include neighbor normalization. The gate $\sigma_h^n$
uses invariant pair norms and scalar features. Its multiplier is shared
over the angular components at each $j$ and both columns of each Kramers
pair, so it commutes with rotations and $Q_a$.

The integer-rank features are fed into an equivariant tensor convolution,
giving $\bar V_a^{n+1}$. The updated spinors then supply a tensor correction
$\mathcal R_{\mathrm{layer}}^n$ through conjugated contractions and angular
projection, as shown in Fig.~\ref{fig:espinor-architecture}(c).
With an invariant channel gate $s_a^n$, the integer update is
\begin{equation}
V_a^{n+1}=\bar V_a^{n+1}
+s_a^n\odot\mathcal R_{\mathrm{layer}}^n(h_a^{n+1},\bar V_a^{n+1}).
\label{eq:tensor-layer-correction}
\end{equation}
\end{subequations}
Here $\odot$ denotes channelwise multiplication. Atoms without spinors
receive only $\bar V_a^{n+1}$. Both $V_a^{n+1}$ and $h_a^{n+1}$ enter the
next layer, so contraction informs the integer features without discarding
the spinor features. The contraction formulas, gate inputs, and construction
of $\mathcal R_{\mathrm{layer}}^n$ are detailed in the Supplemental
Material~\cite{supplementalMaterial} (Sec.~\ref{supp-sec:angular-products}).

The channel maps act identically on all eigenbasis columns. Together with
the angular projections and invariant gates shared within each Kramers
pair, this guarantees the spinor transformation law in
Eq.~\eqref{eq:spinor-initial-law} at every layer.
The contraction over $u$ in Eq.~\eqref{eq:three-couplings} removes the
eigenbasis dependence, and angular projection gives gauge-invariant
integer-rank tensors with the required rotation laws, as proved in
\apxref{app:spinor-symmetry}.

\subsection{Tensor readout}
\label{sec:tensor-readout}
\label{sec:symmetry-bias}

To predict magnetic tensors, the readout combines the final integer-rank
features with gauge-invariant spinor contractions. After
$N_{\mathrm{lay}}$ interaction layers, we omit the layer index from $h_a$
and $V_a$ and write the metal-site readout in
Fig.~\ref{fig:espinor-architecture}(a) as
\begin{equation}
\hat y_a^{(\ell)}=W_{\mathrm{out}}^{(\ell)}V_a^{(\ell)}+\mathcal R_{\mathrm{out}}^{(\ell)}(h_a,V_a).
\label{eq:spinor-readout}
\end{equation}
The equivariant map $W_{\mathrm{out}}^{(\ell)}$ combines channels of rank
$\ell$, while $\mathcal R_{\mathrm{out}}$ uses conjugated spinor contractions
projected onto the same rank.

Under the initialization assumptions in Secs.~\ref{sec:initialization}
and~\ref{sec:gauge}, including separated Kramers pairs, both contributions
are gauge-invariant tensors of rank $\ell$. This guarantees that the
prediction is independent of the eigenbasis and obeys the required
$\SO{3}$ rotation law,
\begin{equation}
\hat y_a^{(\ell)}(g\cdot G)=D^{(\ell)}(g)\hat y_a^{(\ell)}(G).
\label{eq:prediction-law}
\end{equation}
We prove this result in \apxref{app:spinor-symmetry} and report
numerical equivariance tests in the Supplemental
Material~\cite{supplementalMaterial} (Sec.~\ref{supp-sec:numerical-checks}).

For the Co(II) benchmark, we convert the rank-$2$ output to the symmetric,
traceless Cartesian ZFS tensor $\widehat{\mathbf D}$. For the Dy(III)
crystal-field benchmark, we convert
outputs of ranks $2$, $4$, and $6$ to Stevens coefficients
$\widehat B_q^{(\ell)}$ using the fixed basis maps given in the
Supplemental Material~\cite{supplementalMaterial}. The physical definitions
of these targets follow in Secs.~\ref{sec:cobalt} and~\ref{sec:dysprosium}.

Propagating spinors before contraction also introduces coefficient sharing.
In a product $hh^\dagger$, the learned update of $h$ determines both
factors, linking the coefficients of the resulting tensor interactions,
as derived in the Supplemental Material~\cite{supplementalMaterial}
(Sec.~\ref{supp-sec:spinor-factorization}).
To assess how the architecture uses electronic information, we compare
eSpinor with NequIP and its Hamiltonian-informed ablation NequIP+H.
NequIP constructs integer-rank features from atomic species and geometry.
NequIP+H also injects gauge-invariant contractions of Hamiltonian
eigenvectors at each layer and propagates the resulting integer features.
eSpinor additionally updates and propagates spinor features before
contraction. The Hamiltonian-informed models use comparable parameter
budgets. All models are trained and selected using static magnetic-property
errors. Their implementation and training procedures are specified in the
Supplemental Material~\cite{supplementalMaterial}
(Sec.~\ref{supp-sec:training-details}).

\section{Zero-field splitting in cobalt complexes}
\label{sec:cobalt}

Zero-field splitting (ZFS) is the separation of spin energy levels without
an applied magnetic field. In Co(II) complexes, SOC couples the spin to
the ligand-dependent orbital structure, producing magnetic
anisotropy~\cite{boca2004zfs,mariano2024compass}. Easy-axis anisotropy favors
opposite orientations of the magnetic moment along a molecular axis and
can hinder transitions between them, supporting the slow magnetic
relaxation sought in SMMs~\cite{craig2015sim3d}.
Changing the ligands can alter both the magnitude and character of this
anisotropy. Predicting ZFS across ligand environments therefore helps
identify promising molecular magnets.

\begin{table*}[t]
  \renewcommand{\arraystretch}{1.2}
  \centering
  \caption{Cobalt prediction MAEs (cm$^{-1}$) on the splits in
  Sec.~\ref{sec:cobalt}. The tensor error averages all 9 Cartesian
  entries. Parameter counts are ridge coefficients for eqSNAP or
  trainable neural parameters.
  The eSpinor variants use point-charge (pc) or angular-overlap (AOM)
  initialization.}
  \label{tab:cobalt-tensors}
  \begin{tabular*}{\textwidth}{@{\extracolsep{\fill}}lrcccccc@{}}
    \toprule
    & & \multicolumn{3}{c}{In distribution ($n=862$)}
      & \multicolumn{3}{c}{Unseen ligands ($n=473$)} \\
    \cmidrule(lr){3-5}\cmidrule(lr){6-8}
    Model & Parameters & $\mathbf D$ tensor & $D$ & $|E|$ & $\mathbf D$ tensor & $D$ & $|E|$ \\
    \midrule
    eqSNAP & 1,870 & 3.674 & 16.832 & 1.729 & 2.989 & 12.870 & 1.670 \\
    GM-NN & 943,874 & 2.380 & 12.090 & 1.417 & 2.988 & 14.710 & 1.666 \\
    NequIP & 122,120 & 1.193 & 6.645 & 0.835 & 1.651 & 7.611 & 1.134 \\
    NequIP+H & 144,923 & 1.085 & 5.925 & 0.825 & 1.626 & 8.619 & 1.211 \\
    eSpinor (pc) & 149,332 & \secondbest{1.004} & \secondbest{5.505} & \secondbest{0.709}
      & \secondbest{1.370} & \secondbest{6.893} & \secondbest{1.061} \\
    eSpinor (AOM) & 150,679 & \best{0.849} & \best{4.838} & \best{0.641}
      & \best{1.264} & \best{6.365} & \best{0.875} \\
    \bottomrule
  \end{tabular*}
  \par\smallskip
  \begin{minipage}{\textwidth}
    \footnotesize\raggedright *Best predictive results are bold and second best are underlined
    throughout the result tables.
  \end{minipage}
\end{table*}

COMPASS was developed to identify Co(II) complexes with large magnetic
anisotropy for SMM design~\cite{mariano2024compass,mariano2024compassdata}.
We use its 4-coordinate complexes with CASSCF-SO reference targets to
test ZFS prediction and generalization to new ligands. All models share
6892 training, 862 validation, 862 in-distribution test, and 473
out-of-distribution (OOD) test structures. The OOD complexes contain
only ligands absent from training. Data preprocessing and split construction
are specified in the Supplemental Material~\cite{supplementalMaterial}
(Sec.~\ref{supp-sec:protocol}).

NequIP+H and eSpinor use AOM initialization, with point-charge eSpinor
as an additional comparison. We also evaluate eqSNAP, which uses linear
regression on equivariant neighbor-density descriptors~\cite{nguyen2022eqsnap},
and GM-NN, which learns invariant coefficients of geometric tensor
features~\cite{zaverkin2022gmnn}.

\subsection{Effective spin Hamiltonian}
\label{sec:cobalt-task}

The reference calculations represent the lowest 2 Kramers pairs by an
effective spin $S=3/2$~\cite{mariano2024compass,chibotaru2012pseudospin}.
We predict the real, symmetric, traceless tensor $\mathbf D$ in the
ZFS Hamiltonian
\begin{equation}
  H_{\mathrm{ZFS}}=\sum_{\alpha,\beta=x,y,z}
  D_{\alpha\beta}S_\alpha S_\beta,
  \label{eq:cobalt-zfs-hamiltonian}
\end{equation}
where $S_\alpha$ are dimensionless spin operators. The tensor's 5 independent
components specify the strength and orientation of the anisotropy.

In the principal-axis frame of $\mathbf D$, $D=3D_{zz}/2$ describes axial
anisotropy and $E=(D_{xx}-D_{yy})/2$ measures the difference in anisotropy
between the $x$ and $y$ directions~\cite{boca2004zfs}.
For $E=0$, negative $D$ gives easy-axis
anisotropy, whereas positive $D$ gives easy-plane anisotropy. Both $D$ and
$|E|$ matter for assessing SMM candidates~\cite{mariano2024compass}, so we
evaluate their errors alongside the full tensor error, which is also
sensitive to the orientation of the principal axes.
We specify the principal-axis convention and MAE definitions in the
Supplemental Material~\cite{supplementalMaterial} (Sec.~\ref{supp-sec:cobalt-details}).

\subsection{Prediction accuracy and OOD ligand generalization}
\label{sec:cobalt-results}

eSpinor (AOM) gives the lowest tensor, $D$, and $|E|$ errors on both test
sets (Table~\ref{tab:cobalt-tensors}). Relative to NequIP, it reduces the
tensor MAE by $29\%$ in distribution and $23\%$ for unseen ligands.
Relative to NequIP+H, the reduction is approximately $22\%$ on both tests.
NequIP+H lowers the in-distribution tensor error relative to NequIP but
offers little improvement for unseen ligands, where its $D$ and $|E|$
errors exceed NequIP's. Thus, NequIP+H's use of Hamiltonian-derived
integer features does not consistently improve ligand generalization
over NequIP in this comparison.

Within eSpinor, AOM lowers all reported errors compared with point-charge
initialization. This gain is consistent with AOM's description of
directional metal--ligand orbital interactions, which change ligand-field
splittings and hence the anisotropy produced by
SOC~\cite{schaffer1965aom,mariano2024compass}.
However, point-charge eSpinor also gives lower errors than
AOM-initialized NequIP+H for all 3 metrics on both test sets. The advantage
therefore cannot be attributed to AOM alone.

To assess the relevance for screening, we compare candidate rankings and
individual prediction errors for NequIP, NequIP+H, and eSpinor (AOM).
On unseen ligands, eSpinor improves the ordering of complexes by $D$,
raising the Spearman rank correlation between predicted and reference
values from $0.757$ for NequIP and $0.675$ for NequIP+H to $0.787$.
It also lowers the 95th percentile of the per-complex tensor error
from $4.73$ and $5.88$ to $3.41$~cm$^{-1}$, respectively.
The improvement extends to overall candidate ordering and larger errors.
For shortlists ranked by most negative $D$, the model with the largest overlap
with the reference shortlist depends on how many complexes are selected.
These diagnostics and the comparisons in Fig.~\ref{supp-fig:cobalt-screening}
are presented in the Supplemental Material~\cite{supplementalMaterial}.

These comparisons support ligand generalization within 4-coordinate
Co(II) complexes. Transfer to other ions or
coordination numbers and prediction of SMM relaxation times require further testing.

\FloatBarrier

\section{Magnetic excitation spectra in dysprosium complexes}
\label{sec:dy-spectra}

In Dy(III) SMMs, ligand-field design seeks large separations between
electronic energy levels while preserving easy-axis magnetic anisotropy.
These features can suppress thermally activated magnetization reversal,
although relaxation also depends on coupling to molecular
vibrations~\cite{frangoulis2026dysmm}. Strong axial donor interactions and
weaker equatorial interactions provide a useful design principle.
However, changing the organic structure beyond the donor atoms can
substantially alter the splittings even within the same coordination
motif~\cite{frangoulis2026dysmm}. Predicting these changes across unseen
ligands could help prioritize molecular candidates for costly
electronic-structure calculations.

\begin{table*}[t]
  \renewcommand{\arraystretch}{1.2}
  \centering
  \caption{Dy25k excitation-energy MAEs (cm$^{-1}$) and screening results
  for 8k training and 2251 unseen-ligand test complexes. Spectrum averages
  all 7 excitations. Screening reports recovery of the 11 reference
  candidates and false positives under the energy criteria in
  Sec.~\ref{sec:dy-spectra-results}.}
  \label{tab:dy-spectra}
  \begin{tabular*}{\textwidth}{@{\extracolsep{\fill}}lrccccc@{}}
    \toprule
    & & \multicolumn{3}{c}{Excitation-energy MAE}
      & \multicolumn{2}{c}{Candidate screening}\\
    \cmidrule(lr){3-5}\cmidrule(lr){6-7}
    Model & Parameters & Spectrum & $\Delta E_{01}$ & $\Delta E_{07}$
      & Recovered & False positives\\
    \midrule
    eqSNAP & 2,317 & 69.19 & 41.41 & 78.45 & 5/11 & 8\\
    GM-NN & 946,728 & 27.11 & 21.94 & 33.25 & \best{11/11} & 3\\
    NequIP & 127,263 & 28.18 & 19.74 & 40.37 & 9/11 & \secondbest{2}\\
    NequIP+H & 165,826 & \secondbest{21.87} & \secondbest{15.26}
      & \secondbest{26.36} & \secondbest{10/11} & 3\\
    eSpinor & 166,086 & \best{17.74} & \best{12.49} & \best{21.47}
      & \best{11/11} & \best{1}\\
    \bottomrule
  \end{tabular*}
\end{table*}

We examine this transfer using the collection of 25,024 generated
complexes~\cite{frangoulis2026dysmm,frangoulis2026dy25kdata}, which we
refer to as Dy25k.
The source study varies the axial
organic ligands $L$ in $[\mathrm{Dy}(\mathrm{H}_2\mathrm O)_5L_2]^{n+}$,
built from a pentagonal bipyramidal template with 5 equatorial water
ligands. It finds that ligand chemistry can tune the crystal-field
splitting over a wide range, including values above 1600~cm$^{-1}$.
This collection tests whether a model can learn the dependence
of the magnetic spectrum on ligand chemistry within a common
coordination motif.

\subsection{Excitation energies and ligand transfer}
\label{sec:dy-spectra-task}

The ligand field splits the Dy(III) ground multiplet with $J=15/2$
into 8 Kramers pairs. We predict the 7 excitation energies
\begin{equation}
\Delta E_{0k}=E_k-E_0,\qquad k=1,\ldots,7,
\label{eq:dy-excitation-energies}
\end{equation}
where $E_k$ denotes the energy of pair $k$ in ascending order.
The first excitation $\Delta E_{01}$ measures the separation from the
ground pair, while $\Delta E_{07}$ gives the total splitting of the
multiplet. Both are used as magnetic screening criteria in the source
study~\cite{frangoulis2026dysmm}. Their rotational invariance allows us
to compare scalar-output versions of NequIP, NequIP+H, eSpinor, GM-NN,
and eqSNAP. This extends the comparison from magnetic tensors to
rotation-invariant spectra, using the released CASSCF-SO energies as
reference targets.

To test ligand generalization at fixed metal ion and coordination number,
we retain 7-coordinate complexes with 2 O-donor or 2 N-donor organic
ligands and hold out entire ligand families. Structures sharing a source
molecule or ligand heavy-atom connectivity remain in the same partition.
Both donor classes occur in training, validation, and testing, so the
test probes new ligand chemistry with familiar coordinating elements.
All 5 models share 8k training complexes, 2244 validation
complexes, and 2251 unseen-ligand test complexes. Preprocessing, scalar
readout, and additional results are given in the Supplemental
Material~\cite{supplementalMaterial} (Secs.~\ref{supp-sec:training-details} and~\ref{supp-sec:dy-spectra-details}).

\subsection{Prediction accuracy and candidate screening}
\label{sec:dy-spectra-results}

eSpinor predicts unseen-ligand excitation spectra more accurately than
all 4 baselines, with the lowest spectrum, first-excitation, and
total-splitting MAEs (Table~\ref{tab:dy-spectra}). Its spectrum MAE is
$35\%$ lower than GM-NN's, $37\%$ lower than NequIP's, and $19\%$ lower
than NequIP+H's, with essentially the same parameter count as the latter.
Compared with GM-NN, eSpinor also has a lower mean absolute error over
the 7 excitation energies for $78\%$ of the test complexes.
The advantage holds separately for O-donor and N-donor complexes, as
shown in the Supplemental Material~\cite{supplementalMaterial}, supporting
transfer to unseen ligands within both coordinating-element classes.

To connect these errors to screening, we apply the source study's
magnetic criteria, $\Delta E_{01}>550$~cm$^{-1}$ and
$\Delta E_{07}>1500$~cm$^{-1}$~\cite{frangoulis2026dysmm}.
Both eSpinor and GM-NN recover all 11 reference candidates, with 1 and
3 false positives, respectively. eSpinor therefore gives a smaller set
for follow-up electronic-structure calculations without missing any
reference candidates. These screening counts complement the energy MAEs,
which distinguish prediction accuracy even when candidate recovery is
complete. The comparison covers the magnetic
energy criteria, with synthetic accessibility and relaxation times left
for further assessment. Recovery of the highest-energy shortlists also
depends on their size, as shown in the Supplemental
Material~\cite{supplementalMaterial} (Fig.~\ref{supp-fig:dy-spectra-learning}).

Because excitation energies are invariant under molecular rotation, this
benchmark shows that eSpinor's benefit extends to scalar targets.
Together with the cobalt results, it supports using spinor features to
learn the dependence of both magnetic tensors and excitation spectra on
ligand chemistry.

\FloatBarrier

\section{Crystal-field learning and spin dynamics in a dysprosium complex}
\label{sec:dysprosium}

We next consider the Dy(III) complex [DyL$^1$], a paramagnetically
shifted (PARASHIFT) MRI probe~\cite{thompson2026parashift}.
In this type of probe, magnetic anisotropy shifts the nuclear magnetic
resonances of protons on the ligands away from the water and lipid
background, allowing the probe's signal to be detected
separately~\cite{harvey2013parashift}.
Nuclear spin relaxation governs the recovery of magnetization and the
width of these resonances, linking probe sensitivity to electron-spin
dynamics and molecular motion~\cite{thompson2026parashift}.

In solution, molecular motion changes the ligand field and drives
fluctuations of the electronic magnetic moment that contribute to
ligand-proton relaxation. The original PARASHIFT
study~\cite{thompson2026parashift} combined molecular dynamics with
CASSCF-SO calculations along a trajectory to resolve nonexponential
electron-spin dynamics and predict longitudinal nuclear relaxation rates
in agreement with experiment. This system therefore provides a test of
whether magnetic tensors learned at individual geometries also reproduce
spin dynamics and nuclear relaxation.

The released PARASHIFT data provide reference crystal-field coefficients
at 1~fs intervals~\cite{chilton2025parashiftdata}. We use nested training
sets of 500, 1k, 2k, 4k, and 8k geometries, with a common 1k-frame
validation set and 39k-frame test trajectory. These sets occupy separate
time intervals to avoid interleaving adjacent frames. We compare NequIP,
NequIP+H, eSpinor, and eqSNAP, using point-charge initialization for the
Hamiltonian-informed models. Preprocessing and exact splits are given in
the Supplemental Material~\cite{supplementalMaterial}
(Sec.~\ref{supp-sec:protocol}). We evaluate crystal-field coefficients,
electron-spin correlations, and nuclear relaxation rates against the
CASSCF-SO reference, first in water and then after transfer to methanol
without retraining.

\subsection{Hamiltonian and spin propagation}
\label{sec:dysprosium-physics}

The crystal field splits and mixes the $16$ states of the Dy(III) ground
multiplet with total angular momentum $J=15/2$. Within this space, the
Stevens-operator convention~\cite{stevens1952operators} gives
\begin{equation}
H_{\mathrm{CF}}(t)=\sum_{\ell=2,4,6}\sum_{q=-\ell}^{\ell}
B_q^{(\ell)}(t)O_q^{(\ell)}(\mathbf J).
\label{eq:dy-crystal-field}
\end{equation}
The prediction targets are the $27$ real coefficients $B_q^{(\ell)}$,
expressed in cm$^{-1}$ with Stevens factors absorbed. The operators
$O_q^{(\ell)}$ are dimensionless.

Along the molecular trajectory, the changing coefficients drive spin
evolution under $H_{\mathrm{eff}}(t)=H_{\mathrm{CF}}(t)+H_{\mathrm Z}$,
where $H_{\mathrm Z}$ is the Zeeman interaction with the applied field.
Replacing the reference coefficients by predictions changes both the
electronic energy levels and their evolution.
We derive the propagation and relaxation equations in
\apxref{app:spin-relaxation} and specify their numerical evaluation
in the Supplemental Material~\cite{supplementalMaterial}
(Sec.~\ref{supp-sec:spin-dynamics-protocol}).

\subsection{Crystal-field prediction}
\label{sec:dysprosium-learning}

Figure~\ref{supp-fig:parashift-lc} of the Supplemental
Material~\cite{supplementalMaterial} compares the coefficient MAEs, averaged
over components within each rank and over test frames. The neural models
outperform the tested eqSNAP configuration, with similar errors at 8k
and no consistent neural-model ranking across ranks and training-set sizes.

Similar static tensor errors can nevertheless produce different dynamics.
The coefficient errors enter through different Stevens operators and act
throughout the propagation described in \apxref{app:spin-relaxation},
where their signs and temporal ordering matter.
Rankwise MAEs average away this information. We therefore compare the
propagated observables directly.

\subsection{Electron-spin correlations}
\label{sec:dysprosium-dynamics}

The longitudinal electron correlation measures how much angular momentum
along the field $B_0\hat{\mathbf n}$ remains correlated after a lag $\tau$.
With $J_\parallel=\hat{\mathbf n}\cdot\mathbf J$ and propagator
$P(\tau;t_0)$, we compute
\begin{equation}
g_{zz}(\tau;B_0)=\left\langle\operatorname{Tr}
[\rho_0J_\parallel P^\dagger J_\parallel P]\right\rangle_{t_0,\hat{\mathbf n}}.
\label{eq:dy-correlation}
\end{equation}
Here $\rho_0$ is the initial thermal state at $300$~K, and the brackets
average starting times and field orientations. We compare the real curves
through relative $L^2$ errors over $0$--$499$~fs, without fitted tails.
Errors are averaged equally over
$B_0=1$, $1.5$, $2$, $3$, $10$, and $20$~T
(see the Supplemental Material~\cite{supplementalMaterial},
Sec.~\ref{supp-sec:curve-errors}).

\begin{figure*}[!t]
\centering
\includegraphics[width=\textwidth]{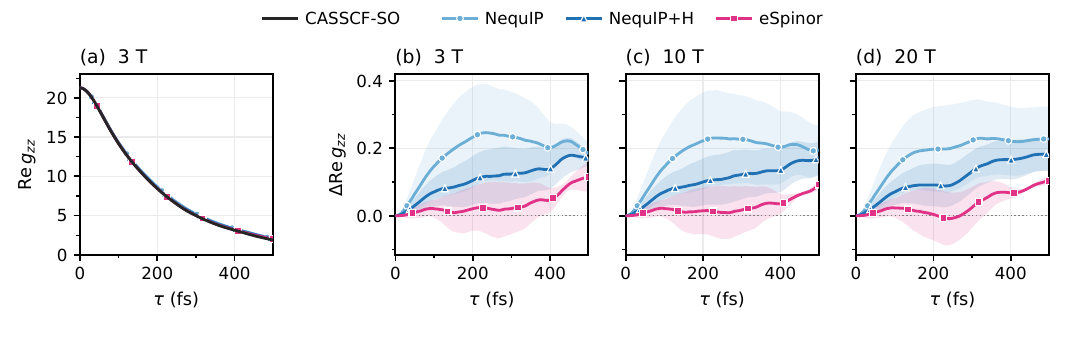}
\caption{Electron-spin correlations from neural models trained on 8k samples.
(a) Raw $\operatorname{Re}g_{zz}$ at $3$~T, with CASSCF-SO as reference.
(b)--(d) Signed model-minus-reference residuals at $3$, $10$, and $20$~T
on a common scale. Dotted lines mark zero error, and shaded bands show
pointwise standard deviations. Curves retain lags $0$--$499$~fs without
a fitted tail. Results for eqSNAP are included in Table~\ref{tab:parashift-dynamics}
and in Fig.~\ref{supp-fig:observable-learning} of the Supplemental
Material~\cite{supplementalMaterial}.}
\label{fig:electron-dynamics}
\end{figure*}

\begin{figure*}[!t]
\centering
\includegraphics[width=\textwidth]{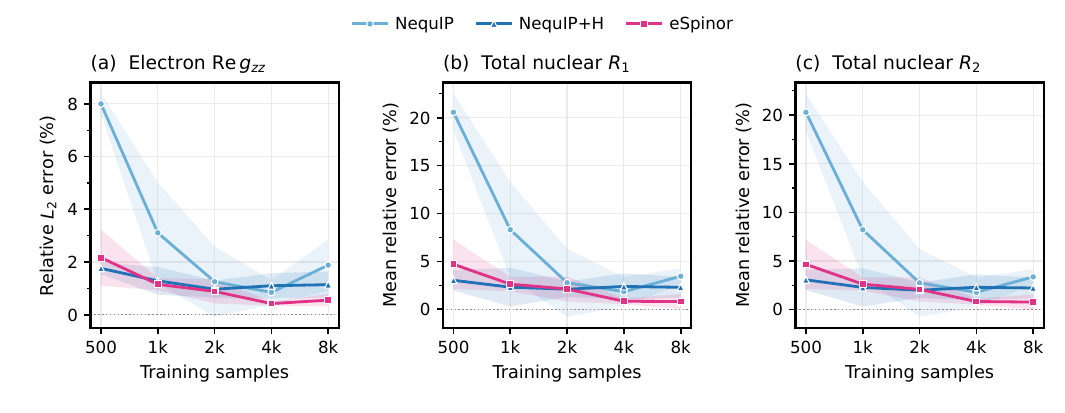}
\caption{Downstream learning curves for NequIP, NequIP+H, and eSpinor.
(a) Real electron-correlation relative $L^2$ error averaged over all 6 fields.
(b), (c) Total nuclear $R_{1N}$ and $R_{2N}$ errors over H3/H4/H5 and
$1$, $1.5$, $2$, and $3$~T.
All errors are relative to CASSCF-SO on the common 39k test frames.
Shaded bands show standard deviations.
Figure~\ref{supp-fig:observable-learning} of the Supplemental
Material~\cite{supplementalMaterial} includes eqSNAP.}
\label{fig:downstream-learning}
\end{figure*}

eSpinor gives the lowest mean
real-correlation error from 1k through 8k, whereas NequIP+H leads at 500
samples [Fig.~\ref{fig:downstream-learning}(a)]. At 8k, eSpinor reduces the
error from NequIP+H's $1.140\%$ to $0.553\%$
(Table~\ref{tab:parashift-dynamics}), although neither model is trained on
dynamics. Figure~\ref{fig:electron-dynamics} resolves this difference
through signed residuals. The advantage persists after normalizing each
curve by its initial value and across separate trajectory segments,
so it extends to the decay shape
(see the Supplemental Material~\cite{supplementalMaterial}).

The advantage also survives using the reference initial thermal state
for every model at $3$ and $20$~T. Individual-window and orientation errors
are comparable to NequIP+H's, but eSpinor has a smaller ensemble-mean
signed error, indicating less systematic error across the sampled windows
and orientations. We define this diagnostic in the Supplemental
Material~\cite{supplementalMaterial}
(Table~\ref{supp-tab:dy-propagation-diagnostics}).

The mean real-correlation error increases from 4k to 8k for all 3 neural
models, even as coefficient MAEs decrease. Training and checkpoint
selection optimize tensor accuracy, while propagated correlations also
depend on how coefficient errors combine across tensor components and
time. Identifying which changes cause the observed increase requires
further investigation. Imaginary correlations and fitted decay parameters
also give different rankings (see the Supplemental
Material~\cite{supplementalMaterial}). These results reinforce the need
to evaluate the intended physical observable directly.

\subsection{Nuclear relaxation rates}
\label{sec:dysprosium-rates}

Nuclear relaxation connects spin dynamics to MRI signal recovery and
linewidth. The longitudinal rate $R_{1N}=T_{1N}^{-1}$ describes recovery
of magnetization along the field, while $R_{2N}=T_{2N}^{-1}$ describes
transverse signal decay~\cite{thompson2026parashift}.

The dipolar contribution combines the full electronic correlation tensor
with the changing Dy--proton geometry through the correlation $G_p(\tau)$
for proton $p$. Molecular tumbling also modulates susceptibility-dependent
shielding, producing Curie relaxation. Following
PARASHIFT~\cite{thompson2026parashift}, we combine these mechanisms as
\begin{equation}
R_{aN,p}=R_{aN,p}^{\mathrm{dip}}+R_{aN,p}^{\mathrm{Curie}},\qquad a=1,2.
\label{eq:dy-rate-components}
\end{equation}
We derive the correlation integrals in \apxref{app:spin-relaxation}.
The fitted tails and ground-multiplet approximation used for model
susceptibilities are detailed in the Supplemental
Material~\cite{supplementalMaterial} (Sec.~\ref{supp-sec:rate-construction}).

We report mean absolute relative errors in the proton-group rates for
H3/H4/H5 at $1$, $1.5$, $2$, and $3$~T using the definitions in the
Supplemental Material~\cite{supplementalMaterial}. We also compare
$G_h=\frac13\sum_{p\in h}G_p$ before fitting and integration through
relative $L^2$ errors averaged over these groups and all 6 fields.

\begin{table}[!b]
\centering\small
\setlength{\tabcolsep}{2pt}
\renewcommand{\arraystretch}{1.2}
\caption{Dy downstream errors (\%) relative to CASSCF-SO for 8k training
samples and 39k test frames. Metrics are defined in
Secs.~\ref{sec:dysprosium-dynamics} and \ref{sec:dysprosium-rates}.}
\label{tab:parashift-dynamics}
\begin{tabular*}{\columnwidth}{@{\extracolsep{\fill}}lrrrrr@{}}
\toprule
Model & Parameters & \shortstack{Re $g_{zz}$\\1--20 T}
  & \shortstack{Complex $G_h$\\1--20 T}
  & \shortstack{$R_{1N}$\\1--3 T} & \shortstack{$R_{2N}$\\1--3 T} \\
\midrule
eqSNAP & 1,650 & 10.935 & 11.735 & 22.687 & 22.386 \\
NequIP & 122,080 & 1.882 & 2.121 & 3.448 & 3.361 \\
NequIP+H & 159,116 & \secondbest{1.140} & \secondbest{1.246}
  & \secondbest{2.294} & \secondbest{2.220} \\
eSpinor & 134,459 & \best{0.553} & \best{0.748}
  & \best{0.809} & \best{0.774} \\
\bottomrule
\end{tabular*}
\end{table}

eSpinor gives the lowest mean $R_{1N}$ and $R_{2N}$ errors at 4k and 8k
(Table~\ref{tab:parashift-dynamics}). At 8k, the absolute rate MAEs
are $0.51$ and $0.49$~s$^{-1}$, compared with $1.40$ and
$1.38$~s$^{-1}$ for NequIP+H. Its advantage also holds for
$G_h$ and the dipolar rates, connecting the improvement in spin dynamics
to the relaxation mechanism that dominates in this field range.
The ranking changes at smaller
training sets, where eqSNAP leads at 500 and NequIP+H at 1k and 2k
(Figs.~\ref{supp-fig:observable-learning} and~\ref{supp-fig:field-mechanisms}
in the Supplemental Material~\cite{supplementalMaterial}).

At $20$~T, the susceptibility-dependent Curie mechanism supplies about
$93\%$ of the reference $R_{2N}$, and NequIP gives the lowest total-rate
error, as shown in Fig.~\ref{supp-fig:field-mechanisms} of the Supplemental
Material~\cite{supplementalMaterial}. The change in ranking follows
the shift from dipolar to Curie relaxation, which probes a different
electronic response.

Correlation and rate comparisons remain complementary because integral
and fitted-tail errors can compensate. In the Supplemental
Material~\cite{supplementalMaterial}, we examine this cancellation and show
that varying the tail cutoff and reference
susceptibility treatment preserves the 8k low-field ranking.

\subsection{Transfer from water to methanol}
\label{sec:dysprosium-transfer}

We evaluate the 8k water-trained models on the methanol (MeOH) trajectory
without retraining~\cite{thompson2026parashift}. Inputs include nearby
solvent atoms, so this probes transfer to the geometries and local environments
sampled in another solvent (see the Supplemental Material~\cite{supplementalMaterial},
Sec.~\ref{supp-sec:meoh-transfer}).

\begin{table}[!htbp]
\centering\small
\setlength{\tabcolsep}{2pt}
\renewcommand{\arraystretch}{1.2}
\caption{Transfer errors (\%) relative to the MeOH CASSCF-SO reference
using models trained on 8k water geometries.
The trajectory and metrics are specified in the Supplemental
Material~\cite{supplementalMaterial}.}
\label{tab:parashift-transfer}
\begin{tabular*}{\columnwidth}{@{\extracolsep{\fill}}lrrrrr@{}}
\toprule
Model & \shortstack{Re $g_{zz}$\\1--20 T}
  & \shortstack{Complex $G_h$\\1--20 T}
  & \shortstack{$R_{1N}$\\1--3 T} & \shortstack{$R_{2N}$\\1--3 T}
  & $\boldsymbol\chi T$ \\
\midrule
eqSNAP & 9.289 & 8.402 & 11.304 & 11.219 & 19.851 \\
NequIP & 2.427 & 2.427 & 3.209 & 3.186 & 5.423 \\
NequIP+H & \secondbest{2.110} & \secondbest{1.982}
  & \secondbest{2.296} & \secondbest{2.281} & \best{5.221} \\
eSpinor & \best{1.240} & \best{0.979}
  & \best{0.959} & \best{0.954} & \secondbest{5.263} \\
\bottomrule
\end{tabular*}
\end{table}

Table~\ref{tab:parashift-transfer} indicates that eSpinor achieves the lowest electron-spin and proton-correlation errors, while both nuclear-rate errors stay under $1\%$. This shows that eSpinor’s dynamical benefit carries over to a different solvent trajectory, despite being trained only on data from water. In contrast, susceptibility errors for $\boldsymbol\chi T$ are comparable across the neural models.

\section{Discussion}
\label{sec:discussion}

\subsection{OOD ligand generalization}
\label{sec:discussion-ligands}

The ligand tests in Secs.~\ref{sec:cobalt} and~\ref{sec:dy-spectra}
hold the metal ion and coordination number fixed within each benchmark.
They therefore probe how well the models learn changes in anisotropy
and excitation spectra caused by unfamiliar ligand chemistry.
For Co(II), better ordering of axial anisotropy and a lower
95th-percentile tensor error connect eSpinor's improvement to candidate
screening. For Dy(III) SMM candidates, the excitation spectra provide
criteria for selecting complexes with large magnetic splittings.
These results support prioritizing unfamiliar ligands for
electronic-structure calculations in searches for molecular
magnets~\cite{frangoulis2025genetic}. Prospective screening and transfer
to other ions and coordination numbers require further testing.

The initializer comparison also has a physical interpretation. AOM describes
directional metal--ligand interactions that alter orbital energies and
the anisotropy produced by SOC~\cite{schaffer1965aom,mariano2024compass}.
Its improvement over point-charge initialization is consistent with this
role. Yet point-charge eSpinor also outperforms AOM-initialized NequIP+H.
A suitable ligand-field description helps, and eSpinor's advantage over
NequIP+H persists with either initializer.

\subsection{From coefficients to magnetic observables}
\label{sec:discussion-observables}

For the Dy(III) MRI probe, the clearest distinction appears in the
dynamical consequences of the predicted tensors. At 8k, similar
crystal-field MAEs lead to different
electron-spin correlations and nuclear relaxation rates
(Sec.~\ref{sec:dysprosium}). Propagation weights errors by their action
on electronic states, and the correlation averages over molecular
motion and field orientations. The diagnostic with a common initial
thermal state identifies a smaller systematic residual for eSpinor in
the sampled ensemble, despite comparable individual-sample errors.
The diagnostic thus locates the gain at the ensemble level. Further
work should examine which tensor components and temporal error
correlations account for this difference.

The relevant accuracy criterion also changes with the relaxation mechanism.
Dipolar rates depend on spin correlations and proton motion, whereas Curie
rates depend quadratically on anisotropic shielding under molecular
tumbling, as detailed in \apxref{app:spin-relaxation} and the
Supplemental Material~\cite{supplementalMaterial}. This explains why improved
dipolar dynamics can coexist with a different high-field ranking, and why
the full susceptibility error alone does not determine Curie-rate accuracy.
Evaluating both correlations and integrated rates therefore gives a more
complete assessment of a surrogate for relaxation calculations.

The MeOH test extends the comparison beyond the water trajectory without
additional electronic-structure training data. The inputs include nearby
solvent atoms, so this result supports transferring the learned mapping
to geometries and local environments sampled under another solvent condition.

For 8k eSpinor models, exporting 39k predictions averaged about 12~min on
1 NVIDIA H200 GPU, or 18~ms per geometry including setup and I/O.
For context, 2 approximate CASSCF-SO timing jobs took 1.4--2.3~h per
geometry on 72 NVIDIA Grace CPU cores on DeltaAI.
The Supplemental Material~\cite{supplementalMaterial} specifies their
different electronic-structure settings and the timing scope.
This low inference cost, together with the demonstrated water-to-MeOH
transfer, could enable longer and more densely sampled trajectories for
spin-dynamics calculations using a limited reference set.

\subsection{SU(2) inductive bias}
\label{sec:discussion-bias}

We compare eSpinor with NequIP+H to assess the benefit of the spinor
architecture beyond supplying Hamiltonian-derived information.
NequIP+H contracts weighted Hamiltonian eigenvectors into gauge-invariant
integer-rank features at each layer and propagates these features.
eSpinor additionally updates and propagates spinor features under
half-integer $\SU{2}$ rotation laws, then feeds their gauge-invariant
contractions into the integer-rank features. The comparison therefore
tests how electronic information is represented and processed within
the network.

The Dy(III) SMM candidate benchmark makes this distinction particularly
clear. Excitation energies are rotationally invariant scalars, and both
architectures enforce that output symmetry. Yet eSpinor lowers the
spectrum MAE by $19\%$ with essentially the same parameter count
(Table~\ref{tab:dy-spectra}). The gain therefore supports the usefulness
of its internal representation and interactions for learning how ligand
geometry affects SOC-dependent energy levels, even when the outputs are
scalars.

Spinor propagation constrains how tensor interactions are learned.
When updated spinor features are contracted with their complex conjugates,
the same learned transformation enters both factors, linking the
coefficients of the resulting integer-rank interactions.
Keeping the complete set of these products for each Kramers pair allows
eSpinor's updates to be rewritten entirely in integer-rank tensors, with
identical predictions, as shown in the Supplemental
Material~\cite{supplementalMaterial}.
This exact reformulation preserves eSpinor's coefficient sharing, whereas
NequIP+H uses different tensor updates.
A future comparison with
integer-tensor updates that retain each Kramers pair's information but
learn the interaction coefficients independently would further test the
contribution of coefficient sharing.

\section{Conclusions}
\label{sec:conclusions}

We investigated whether $\SU{2}$-equivariant spinor message passing can
provide a useful inductive bias for predicting molecular magnetic properties.
To test this, we developed eSpinor to propagate Hamiltonian-derived spinors
alongside integer-rank features, with gauge-covariant updates and
gauge-invariant readouts.

For Co(II) SMM candidates, eSpinor (AOM) lowers the
zero-field-splitting tensor MAE relative to NequIP by $29\%$ in distribution
and $23\%$ for unseen ligands. For Dy(III) SMM candidates, it gives the
lowest excitation-energy errors among the 5 models on unseen ligands,
extending the benefit to rotationally invariant scalars. For the Dy(III)
MRI probe, eSpinor improves real electron-spin correlations and total
nuclear relaxation rates at $1$--$3$~T at the larger training sizes.
At 8k, it achieves these gains with crystal-field accuracy similar to
NequIP+H and $15.5\%$ fewer parameters. The dynamical advantage extends
from water to methanol without retraining. Together, these results
support $\SU{2}$-equivariant spinor message passing as a useful inductive
bias for the molecular magnetic properties studied here.

The benefit depends on the observable, with susceptibility and some
high-field relaxation rates favoring integer-rank models. Model selection
should therefore use the intended physical observables, since static
tensor accuracy alone does not determine dynamical accuracy.
These surrogates could support ligand screening and longer trajectory
calculations using limited electronic-structure data. The present
comparisons use electronic-structure references, and their practical
value for SMM screening or MRI-probe design requires further validation.

Our benchmarks are limited to mononuclear complexes because we have not
identified a suitable labeled multinuclear dataset. Generating reference
data for interacting magnetic centers requires treating electron
correlation, SOC, and magnetic exchange together~\cite{bodenstein2022casoci}.
We have formulated parallel transport between metal sites that preserves
covariance under independent local eigenbasis changes, as detailed in the Supplemental
Material~\cite{supplementalMaterial} (Sec.~\ref{supp-sec:multinuclear-transport}).
Future work should test whether this extension improves predictions of
magnetic interactions between ions.

\section*{Acknowledgments}

A.A. acknowledges support in part from the Bren Endowed Chair, the Office of Naval Research through MURI grant N00014-18-1-2624, DARPA ExpMath (HR0011), and the AI2050 Senior Fellow program at Schmidt Sciences.
E.W.S. acknowledges support through the Caltech Summer Undergraduate Research Fellowship.
B.S.K. acknowledges support from the Pritzker AI+Science Fund, DARPA Biological Technologies (HR0011), and the Eddleman Graduate Fellowship.
Y.L. acknowledges support from the Pritzker AI+Science Fund and the Wally Baer and Jeri Weiss Postdoctoral Fellowship.
D.K. acknowledges support from the Pritzker AI+Science Fund and DARPA Biological Technologies (HR0011).
Authors acknowledge the National Center for High-performance Computing (NCHC),
Taiwan, for providing computing resources on Taiwania 2 and Nano 4. This work also used the Delta system at the National Center for Supercomputing Applications through allocation DMR160114 from the Advanced Cyberinfrastructure Coordination Ecosystem: Services \& Support (ACCESS) program, which is supported by National Science Foundation grants \#2138259, \#2138286, \#2138307, \#2137603, and \#2138296.

Claude (Anthropic, Opus 5 and Fable 5) and ChatGPT (OpenAI, GPT-6 Astra)
assisted with literature review and manuscript revision. Codex (OpenAI,
GPT-5.6 Sol) assisted with writing and debugging data-processing, plotting,
and unit-test scripts. The authors directed this assistance, reviewed the
text and code, and take full responsibility for this work.

\section*{Data Availability}

The COMPASS and Dy25k datasets are publicly available from
Zenodo~\cite{mariano2024compassdata,frangoulis2026dy25kdata}, and the
PARASHIFT dataset is available from the University of Manchester
Figshare~\cite{chilton2025parashiftdata}.
The source code and preprocessing scripts will be made publicly available at
\href{https://github.com/AI-QChem/eSpinor}{AI-QChem/eSpinor}
upon publication.

\appendix

\section{Rotation equivariance and eigenbasis independence}
\label{app:spinor-symmetry}

We prove the rotation and eigenbasis transformation laws for the
mononuclear architecture in Sec.~\ref{sec:spinor-method}, assuming
separated Kramers pairs. Linear maps and Clebsch--Gordan couplings are
equivariant. Weights and gate values are invariant under rotations and
eigenbasis changes. Gate multipliers are shared over all angular
components within each irrep and both columns within each Kramers pair.
Other column operations must also commute with every allowed $Q_a$.
Below, $g\in\SO{3}$ denotes a molecular rotation and
$\tilde g\in\SU{2}$ either lift of $g$.

We first establish the Hamiltonian covariance needed for spinor
initialization in Sec.~\ref{sec:initialization}. Both the point-charge
and AOM initializers use the angular expansion
\begin{equation}
H_{\mathrm{CF}}(G)
=\sum_{b,\ell,q}f_{b\ell}
\sqrt{\frac{4\pi}{2\ell+1}}\,
Y_{\ell q}(\widehat{\mathbf R}_b)^*\mathsf C_{\ell q}.
\label{eq:ligand-field-expansion}
\end{equation}
At a given metal site, $\widehat{\mathbf R}_b$ points toward ligand $b$, and
$\mathsf C_{\ell q}$ is a rank-$\ell$ tensor operator acting on the orbital
states of the shell~\cite{varshalovich1988quantum}.
The sums include $\ell=0,2,\ldots,2\lsh$ and $q=-\ell,\ldots,\ell$.
The real scalar strengths $f_{b\ell}$ come from the point-charge or AOM
construction and are unchanged by molecular rotations.
The spherical harmonics and orbital tensor operators transform with the
same rank-$\ell$ rotation matrices. Their matching components in
Eq.~\eqref{eq:ligand-field-expansion} therefore give
$H_{\mathrm{CF}}(g\cdot G)
=D^{(\lsh)}(\tilde g)H_{\mathrm{CF}}(G)D^{(\lsh)}(\tilde g)^\dagger$.
The strengths and operator conventions are specified in the Supplemental
Material~\cite{supplementalMaterial}.

In the coupled basis $|j,m\rangle$, with
$j=\lsh\pm\half$, the joint spin--orbital rotation is
$U_{\mathrm{sh}}(\tilde g)=\bigoplus_{j=\lsh\pm\half}D^{(j)}(\tilde g)$~\cite{littlejohn2021coupling}.
Since $\mathbf L\cdot\mathbf S$ is unchanged under this joint rotation,
the full Hamiltonian obeys
\begin{equation}
H_{\mathrm{init}}(g\cdot G)
=U_{\mathrm{sh}}H_{\mathrm{init}}(G)U_{\mathrm{sh}}^\dagger.
\label{eq:initial-hamiltonian-covariance}
\end{equation}
For this fixed rotation, combining Eq.~\eqref{eq:initial-hamiltonian-covariance}
with Eq.~\eqref{eq:initial-eigensystem} gives
\begin{equation}
\begin{aligned}
H_{\mathrm{init}}(g\cdot G)\,[U_{\mathrm{sh}}\Psi(G)]
&=U_{\mathrm{sh}}[H_{\mathrm{init}}(G)\Psi(G)]\\
&=[U_{\mathrm{sh}}\Psi(G)]E.
\end{aligned}
\label{eq:rotated-eigenbasis}
\end{equation}
Thus $U_{\mathrm{sh}}\Psi(G)$ is an eigenbasis at the rotated geometry with
the same energy levels. A separate diagonalization with the same energy
ordering can return another orthonormal basis, related by a unitary $Q$
satisfying $EQ=QE$.
The ligand field and $\mathbf L\cdot\mathbf S$ are unchanged under time
reversal, while the single-electron time-reversal operator obeys
$\mathcal T^2=-I$. These properties guarantee the Kramers
pairs~\cite{kramers1930rotation,wigner1932zeitumkehr}.
When distinct pairs are separated in energy, $Q=\bigoplus_pQ_p$ with
$Q_p\in\U{2}$. Thus $\Psi(g\cdot G)=U_{\mathrm{sh}}\Psi(G)Q$, giving
the combined rotation and eigenbasis change in
Eq.~\eqref{eq:initial-eigenbasis-covariance}.

If $d$ eigenstates have exactly the same energy, $EQ=QE$ instead permits
arbitrary $\U{d}$ mixing within that eigenspace. For 2 coincident Kramers
pairs, the freedom is therefore $\U{4}$ rather than $\U{2}\times\U{2}$.
The proof extends only when all gate inputs are invariant under this larger
group and column operations commute with it. In particular, weights and
diagonal gate multipliers must be equal across all $d$ columns. Small
nonzero gaps do not enlarge the gauge group, but can make eigenvectors
sensitive to changes in geometry.

We next construct the weighted initialization used in Sec.~\ref{sec:gauge}.
With uniform weights, the complete eigenbasis gives
$\Psi(I_k/k)\Psi^\dagger=I_k/k$, which contains no geometry-dependent
multipoles. Energy-dependent weights allow this contraction to retain
information about the ligand environment. Using the weights defined in the
Supplemental Material~\cite{supplementalMaterial}, write
$W=\operatorname{diag}(w_1,\ldots,w_k)$ and $z=\Psi\sqrt W$.
Rotations preserve the energy levels and therefore leave $W$ unchanged.
Equal weights within each pair imply $Q\sqrt W=\sqrt WQ$.

The invariant gate $S_{\mathrm{init}}=\bigoplus_p s_pI_2$ also commutes
with $Q$. Initialization is $h^0=\mathcal L_{\mathrm{init}}[zS_{\mathrm{init}}]$.
The channel map $\mathcal L_{\mathrm{init}}$ acts at fixed $j$,
identically on angular indices $m$ and eigenbasis columns $u$.
It therefore commutes with right multiplication by $Q$ and obeys
$\mathcal L_{\mathrm{init}}U_{\mathrm{sh}}=\mathcal D\mathcal L_{\mathrm{init}}$,
where $\mathcal D$ rotates all output channels. Consequently,
\begin{equation}
\begin{aligned}
z'&=U_{\mathrm{sh}}\Psi Q\sqrt W
=U_{\mathrm{sh}}zQ,\\
h^{0\prime}
&=\mathcal L_{\mathrm{init}}[U_{\mathrm{sh}}zS_{\mathrm{init}}Q]
=\mathcal D h^0Q.
\end{aligned}
\label{eq:weighted-initial-covariance}
\end{equation}
The prime denotes features at the rotated geometry, including their
possible change of eigenbasis. For each rank $j$, this is the feature law
in Eq.~\eqref{eq:spinor-initial-law}. Gauge covariance requires equal
weights within each pair, but does not require the Boltzmann form.

The message-passing operations in Sec.~\ref{sec:couplings} must preserve
Eq.~\eqref{eq:spinor-initial-law}. The brackets in that section denote
Clebsch--Gordan projection. For the mixed product in
Eq.~\eqref{eq:three-couplings}, let $P_{j_o}$ project a rank-$j_i$
spinor $h$ and a rank-$\ell$ integer tensor $V$ onto output rank $j_o$.
For one pair of feature channels, its components are
\begin{equation}
[h^{(j_i)}\otimes V^{(\ell)}]_{j_o,m_o,u}
=\sum_{m_i,q}C^{j_om_o}_{j_im_i,\ell q}
h^{(j_i)}_{m_i,u}V^{(\ell)}_q.
\label{eq:spinor-tensor-components}
\end{equation}
Here $C^{j_om_o}_{j_im_i,\ell q}$ is a Clebsch--Gordan coefficient,
and $j_o=|j_i-\ell|,\ldots,j_i+\ell$ in steps of $1$~\cite{varshalovich1988quantum}.
The projection couples the angular indices $m_i$ and $q$ while leaving
the eigenbasis index $u$ unchanged. Taking
$V^{(\ell)}=Y^{(\ell)}(\hat{\mathbf r}_{ab})$
gives the neighbor coupling in Eq.~\eqref{eq:spinor-residual-update}.
For $h'=D^{(j_i)}hQ$ and $V'=D^{(\ell)}V$, its intertwining property gives
\begin{equation}
\begin{aligned}
P_{j_o}(D^{(j_i)}\otimes D^{(\ell)})&=D^{(j_o)}P_{j_o},\\
P_{j_o}[h'\otimes V']&=D^{(j_o)}P_{j_o}[h\otimes V]Q.
\end{aligned}
\label{eq:cg-intertwiner}
\end{equation}
Invariant scalar weights, equivariant channel maps, gates shared within
each Kramers pair, and residual addition preserve this law. Integer-tensor
products follow the same angular rule without an eigenbasis index.

To form gauge-invariant integer tensors as in
Eq.~\eqref{eq:three-couplings}, let $x$ and $y$ be spinor
features whose eigenbasis columns transform with the same $Q$.
Write $\mathcal D_x$ and $\mathcal D_y$ for their rotation matrices.
Their transformations $x'=\mathcal D_xxQ$ and $y'=\mathcal D_yyQ$ give
\begin{equation}
\begin{aligned}
x'y'^\dagger
&=\mathcal D_xxQQ^\dagger y^\dagger\mathcal D_y^\dagger\\
&=\mathcal D_x(xy^\dagger)\mathcal D_y^\dagger.
\end{aligned}
\label{eq:contraction-proof}
\end{equation}
The gauge cancels because $Q$ is unitary. Products of half-integer
representations and their conjugates decompose into integer ranks, so
angular projections of $xy^\dagger$ give gauge-invariant integer tensors.

For these feature transformation laws, every real-bilinear equivariant
map is a linear combination of Clebsch--Gordan projections with channel
coefficients~\cite{varshalovich1988quantum}. Angular momentum addition gives
integer outputs for equal integer or half-integer types and half-integer
outputs for mixed types. Under a common gauge phase,
$h\mapsto e^{i\phi}h$ and $h^*\mapsto e^{-i\phi}h^*$, whereas $V$ is
unchanged. Thus a mixed product with the same gauge law as $h$ uses
$h\otimes V$, and a gauge-invariant spinor contraction must be conjugated.
For a bilinear column contraction $x_pC_{pp'}y_{p'}^\dagger$ between
Kramers pairs $p$ and $p'$, with $C_{pp'}\in\mathbb C^{2\times2}$,
gauge invariance further requires
\begin{equation}
Q_pC_{pp'}Q_{p'}^\dagger=C_{pp'}
\quad\text{for every allowed }Q.
\label{eq:pair-contraction-condition}
\end{equation}
Independent pair phases force $C_{pp'}=0$ for $p\ne p'$.
For $p=p'$, Schur's lemma gives $C_{pp}=c_pI_2$.
The allowed contractions are therefore angular projections of
$\sum_pc_px_py_p^\dagger$, including scalar weights for each pair.
Together with integer-tensor and mixed products, these give the 3
interaction types in Eq.~\eqref{eq:three-couplings}.

Starting from Eq.~\eqref{eq:weighted-initial-covariance}, induction over
the interaction layers establishes spinor gauge covariance and rotation
equivariance, together with gauge invariance and rotation equivariance of
the integer features. The readout in Sec.~\ref{sec:tensor-readout} uses
these integer tensors and equivariant channel maps, so its predictions
obey Eq.~\eqref{eq:prediction-law}. Changing the lift of a molecular rotation
negates each half-integer factor, and the 2 signs cancel in every
contraction. Predictions are consequently independent of both the initial
eigenbasis and the choice of lift.

\section{From spin propagation to nuclear relaxation}
\label{app:spin-relaxation}

We follow the time-domain treatment of
PARASHIFT~\cite{thompson2026parashift}. The Hamiltonian
$H_{\mathrm{eff}}=H_{\mathrm{CF}}+H_{\mathrm Z}$ in
Sec.~\ref{sec:dysprosium-physics} includes the Zeeman term
$H_{\mathrm Z}=g_J\bar\mu_B B_0J_\parallel$, with $g_J=4/3$ and
$\bar\mu_B=\mu_B/(hc)$ in cm$^{-1}$\,T$^{-1}$. It determines the
propagator $P(\tau;t_0)$ through
\begin{equation}
i\partial_\tau P=2\pi cH_{\mathrm{eff}}(t_0+\tau)P,
\qquad P(0;t_0)=I.
\label{eq:dy-spin-evolution}
\end{equation}
Here $c$ is in cm\,s$^{-1}$ and time is in seconds because the Hamiltonian
is expressed in cm$^{-1}$. Each window starts from its Hamiltonian's
thermal state $\rho_0$. The electronic correlation before averaging is
$g_{\beta\gamma}(\tau)=\operatorname{Tr}
[\rho_0J_\beta P^\dagger J_\gamma P]$, where
$\beta,\gamma\in\{x,y,z\}$. Its longitudinal projection gives
Eq.~\eqref{eq:dy-correlation}.

To obtain nuclear relaxation, we include the dipolar interaction between
the electronic magnetic moment and proton $p$. For a Dy--proton vector
$\mathbf r_p$, its geometry tensor is
$\mathcal D_p=(3\hat{\mathbf r}_p\hat{\mathbf r}_p^T-I)/r_p^3$.
Contracting it with the electronic correlation gives
\begin{equation}
\begin{aligned}
G_p(\tau)=(g_J\mu_B)^2\Big\langle\sum_{\alpha,\beta,\gamma}
&\mathcal D_{p,\alpha\beta}(t_0)
\mathcal D_{p,\alpha\gamma}(t_0+\tau)\\[-2pt]
&\times g_{\beta\gamma}(\tau)\Big\rangle_{t_0,\hat{\mathbf n}}.
\end{aligned}
\label{eq:dy-proton-correlation}
\end{equation}
All 3 indices run over Cartesian components. The contraction precedes
the trajectory and orientation average, retaining the joint fluctuations
of spin and proton geometry. With distances in metres, $G_p$ has units
m$^{-6}$(J/T)$^2$. No product of mean moments is subtracted, following
the source calculation.

The explicit correlation is extended by a real fitted tail, denoted
$\widetilde G_p$. Rotational damping with solvent-dependent correlation
time $\tau_R$ then gives the spectral density
\begin{equation}
\mathcal J_p(\omega)=\int_0^\infty\widetilde G_p(\tau)
e^{-\tau/\tau_R}e^{-i\omega\tau}\,d\tau.
\label{eq:dy-spectral-density}
\end{equation}
At the proton Larmor frequency $\omega_N=\gamma_NB_0$, the dipolar rates
are
\begin{equation}
\begin{aligned}
R_{1N,p}^{\mathrm{dip}}&=\frac{2\kappa}{3}
\operatorname{Re}\mathcal J_p(\omega_N),\\
R_{2N,p}^{\mathrm{dip}}&=\frac{\kappa}{3}
\operatorname{Re}[\mathcal J_p(0)+\mathcal J_p(\omega_N)],
\end{aligned}
\label{eq:dy-dipolar-rates}
\end{equation}
where $\kappa=(\mu_0/4\pi)^2\gamma_N^2$, $\mu_0$ is the vacuum
permeability, and $\gamma_N$ is the proton gyromagnetic ratio. Thus rate
accuracy depends on an integrated correlation, including its extrapolated
tail, rather than on the explicit curve error alone.

The Curie contribution in Eq.~\eqref{eq:dy-rate-components} instead uses
the mean shielding tensor
$\overline{\boldsymbol\sigma}_p
=-\langle\mathcal D_p\boldsymbol\chi_{\mathrm{SI}}\rangle_t/(4\pi)$
in the molecular frame. Here $\boldsymbol\chi_{\mathrm{SI}}$ is the
volume susceptibility per molecule. The shielding anisotropy determines
the Curie rates under molecular tumbling. The susceptibility and Curie
formulas, tail fitting procedure, and sensitivity checks are given in the
Supplemental Material~\cite{supplementalMaterial}.

\nocite{bode2025deltaai,galvan2019openmolcas,gasteiger1980charges,gasteiger2020dimenet,harris2020numpy,hunter2007matplotlib,scheie2021pycrystalfield,vanvleck1932susceptibilities,virtanen2020scipy}

% Supplemental Material for the arXiv version, compiled within main.tex.
\onecolumngrid
\clearpage
\makeatletter
\let\@hangfrom@section\arxiv@hangfrom@section
\let\@sectioncntformat\arxiv@sectioncntformat
\@removefromreset{equation}{section}
\def\theequation@prefix{}
\renewcommand{\subsubsection}{%
  \@startsection{subsubsection}{3}{\z@}%
    {8pt \@plus 2pt \@minus 1pt}{4pt}%
    {\normalfont\small\itshape\centering}%
}
\makeatother
\setcounter{section}{0}
\setcounter{subsection}{0}
\setcounter{subsubsection}{0}
\setcounter{equation}{0}
\setcounter{figure}{0}
\setcounter{table}{0}
\renewcommand{\thesection}{S\arabic{section}}
\renewcommand{\thesubsection}{\Alph{subsection}}
\renewcommand{\thesubsubsection}{\arabic{subsubsection}}
\renewcommand{\theequation}{S\arabic{equation}}
\renewcommand{\thefigure}{S\arabic{figure}}
\renewcommand{\thetable}{S\arabic{table}}
% Separate hyperlink destinations from the main text and appendices.
\renewcommand{\theHsection}{supp.\arabic{section}}
\renewcommand{\theHsubsection}{\theHsection.\arabic{subsection}}
\renewcommand{\theHsubsubsection}{\theHsubsection.\arabic{subsubsection}}
\renewcommand{\theHequation}{supp.\arabic{equation}}
\renewcommand{\theHfigure}{supp.\arabic{figure}}
\renewcommand{\theHtable}{supp.\arabic{table}}
\setlength{\parindent}{1em}
\setlength{\parskip}{0pt}
\setlength{\textfloatsep}{10pt plus 2pt minus 2pt}
\setlength{\intextsep}{8pt plus 2pt minus 2pt}
\setlength{\abovecaptionskip}{6pt}
\setlength{\abovedisplayskip}{5pt plus 1pt minus 1pt}
\setlength{\belowdisplayskip}{5pt plus 1pt minus 1pt}
\setlength{\abovedisplayshortskip}{3pt plus 1pt minus 1pt}
\setlength{\belowdisplayshortskip}{4pt plus 1pt minus 1pt}
\hypertarget{supplemental-material}{}
\pdfbookmark[0]{Supplemental Material}{supplemental-title}
\begin{center}
{\large\bfseries SU(2)-Equivariant Spinor Message Passing for Molecular Magnetism\par
\vspace{0.4em}
Supplemental Material\par}
\vspace{0.7em}
Erh-Wei Sheng\textsuperscript{1,2}, Beom Seok Kang\textsuperscript{3},
Yuchao Lin\textsuperscript{1}, Danish Khan\textsuperscript{1}, and
Anima Anandkumar\textsuperscript{1}\par
\vspace{0.3em}
{\small\itshape
\textsuperscript{1}Division of Engineering and Applied Science, California Institute of Technology,
Pasadena, California 91125, USA\par
\textsuperscript{2}Department of Chemical Engineering, National Tsing Hua University,
Hsinchu 300044, Taiwan\par
\textsuperscript{3}Division of Chemistry and Chemical Engineering, California Institute of Technology,
Pasadena, California 91125, USA\par}
\end{center}
\section{Implementation and symmetry details}
\label{supp-sec:hamiltonian-details}
\label{supp-sec:layer-details}

\subsection{Hamiltonian initialization}
\label{supp-sec:initializer-details}
\label{supp-sec:symmetry-proof}
\label{supp-sec:eigen-derivative}

For Co(II), $(\ell_{\mathrm{sh}},k)=(2,10)$, and for Dy(III), $(3,14)$.
With metal--ligand vectors $\mathbf R_b=R_b\hat{\mathbf R}_b$, the
Condon--Shortley convention~\cite{suppbib-varshalovich1988quantum} gives
\begin{equation}
\begin{aligned}
c_{\ell q}(\hat{\mathbf R})&=\sqrt{\frac{4\pi}{2\ell+1}}Y_{\ell q}(\hat{\mathbf R}),&
(\mathsf C_{\ell q})_{mm'}&=(-1)^m(2\ell_{\mathrm{sh}}+1)
\begin{pmatrix}\ell_{\mathrm{sh}}&\ell&\ell_{\mathrm{sh}}\\0&0&0\end{pmatrix}
\begin{pmatrix}\ell_{\mathrm{sh}}&\ell&\ell_{\mathrm{sh}}\\-m&q&m'\end{pmatrix},\\
H_{\mathrm{CF}}&=\sum_{b,\ell,q}f_{b\ell}c_{\ell q}(\hat{\mathbf R}_b)^*\mathsf C_{\ell q},&
\ell&=0,2,\ldots,2\ell_{\mathrm{sh}}.
\end{aligned}
\label{supp-eq:orbital-construction}
\end{equation}
Here $m,m'$ are orbital projections. The rotation laws are derived in
\apxref{app:spinor-symmetry} of the main text.

Point-charge strengths are $f_{b\ell}^{\mathrm{pc}}=
\kappa Z_b^{\mathrm{eff}}r_\ell^{\mathrm{eff}}/R_b^{\ell+1}$, with
$\kappa=14.3996\times8065.54$~cm$^{-1}$\,\AA.
We use non-hydrogen neighbors within 3~\AA\ and Gasteiger
charges~\cite{suppbib-gasteiger1980charges}. Failed or zero assignments use
$(-.42,-.66,-.28,-.98,-.62,-.52,-.38,-.21,-.17,-.22)$ for
$(\mathrm N,\mathrm O,\mathrm S,\mathrm F,\mathrm{Cl},\mathrm{Br},\mathrm I,
\mathrm P,\mathrm C,\mathrm{Se})$, in units of elementary charge.
The PyCrystalField-based constants~\cite{suppbib-scheie2021pycrystalfield} are
\begin{equation}
\begin{array}{c|rrrrr}
 & \xi/(\mathrm{cm}^{-1})&r_0^{\mathrm{eff}}&r_2^{\mathrm{eff}}&r_4^{\mathrm{eff}}&r_6^{\mathrm{eff}}\\
\hline
\mathrm{Co}^{2+}&-527&1&0.360&0.310&\text{not used}\\
\mathrm{Dy}^{3+}&1900&1&0.849&1.977&10.44
\end{array}
\label{supp-eq:initializer-constants}
\end{equation}
Both Hamiltonian-informed models use these prefactors in \AA$^\ell$ and
distances in \AA. The Dy values omit Bohr-to-\AA\ conversion and shielding,
so they define the feature map rather than calibrated radial moments.

The angular-overlap model learns ligand-axis orbital energies
$e_{b\mu}$~\cite{suppbib-schaffer1965aom}. With
$\Lambda_{\nu\ell}=(\mathsf C_{\ell0})_{\nu\nu}$ and
$\nu=0,\ldots,\ell_{\mathrm{sh}}$, its strengths satisfy
\begin{equation}
f_b^{\mathrm{AOM}}=\Lambda^{-1}e_b,\qquad
e_{b\mu}=\sum_{\nu=\mu}^{2}\operatorname{softplus}(z_{b\nu}),\qquad
z_b=\mathrm{MLP}(V_b^{(0)},\operatorname{emb}(Z_b),R_b).
\label{supp-eq:aom-energies}
\end{equation}
For Co, $\mu=0,1,2$ denotes $\sigma,\pi,\delta$ bonding.
The MLP uses 8 element-embedding components, a width-32 SiLU layer, and
first-convolution scalars. Zero final weights and inverse-softplus biases
initialize $(e_\sigma,e_\pi,e_\delta)\simeq(4000,800,0)$~cm$^{-1}$,
with a $10^{-3}$~cm$^{-1}$ floor.

After adding SOC and diagonalizing in the coupled basis of
\apxref{app:spinor-symmetry}, we use
\begin{equation}
\bar\epsilon_{2p-1}=\bar\epsilon_{2p}=\frac{\epsilon_{2p-1}+\epsilon_{2p}}2,
\qquad w_u=\frac{e^{-\bar\epsilon_u/T_B}}{\sum_v e^{-\bar\epsilon_v/T_B}},
\qquad z=\Psi\sqrt{\operatorname{diag}(w)}.
\label{supp-eq:spinor-weights}
\end{equation}
The learned scale $T_B$ is not a physical temperature. We subtract the lowest
energy and floor $T_B$ at 1~cm$^{-1}$ and $w_u$ at $10^{-12}$ without
renormalizing. The gauge conditions are derived in \apxref{app:spinor-symmetry}.

Eigenvector backpropagation replaces inverse energy gaps by
\begin{equation}
\frac{1}{\epsilon_j-\epsilon_i}\ \longrightarrow\quad
F_{ij}=\frac{\epsilon_j-\epsilon_i}{(\epsilon_j-\epsilon_i)^2+\delta^2},\qquad
F_{ii}=0,\qquad \delta=5~\mathrm{cm}^{-1}.
\label{supp-eq:regularized-gap}
\end{equation}
The resulting Hamiltonian gradient is Hermitized. Only gradients are regularized.
The eigensolver-failure fallback raises precision and adds a diagonal ramp.
The covariance proof and numerical checks use unperturbed Hamiltonians.

\subsection{Feature updates and tensor outputs}
\label{supp-sec:angular-products}
\label{supp-sec:gate-details}
\label{supp-sec:stevens-coordinates}

A fixed unitary map converts real features to the complex spherical basis,
$f_{\mathrm c}^{(\ell)}=U_\ell f_{\mathrm r}^{(\ell)}$, with
$D_{\mathrm c}^{(\ell)}U_\ell=U_\ell D_{\mathrm r}^{(\ell)}$~\cite{suppbib-geiger2022e3nn}.
We use the spinor messages derived in \apxref{app:spinor-symmetry},
divided by the square root of the receiving site's neighbor count. Integer convolutions
use a self-connection and Table~\ref{supp-tab:training-settings}'s normalization.
The conjugated contraction is
\begin{equation}
T_q^{(\ell;j,j')}=\sum_{m,m',u}(-1)^{j'-m'}C^{\ell q}_{jm,j'(-m')}
x_{m,u}^{(j)}(y_{m',u}^{(j')})^*.
\label{supp-eq:spinor-multipoles}
\end{equation}
Its real and imaginary parts in coordinates $U_\ell^\dagger T$ supply
separate real tensor channels.

All gates use SiLU MLPs of hidden width 16, outputs clamped to $[-3,3]$,
and initial multipliers of 1. For pair energies $\mathbf e=(\bar\epsilon_{2p})_p$
and $\nu_p=\sum_{u=2p-1}^{2p}\|\Psi_{\cdot u}\|_2$, initialization uses
\begin{equation}
\mathbf s_{\mathrm{init}}=\operatorname{clip}_{[-3,3]}
\mathrm{MLP}_{\mathrm{init}}
\bigl(\operatorname{std}(\mathbf e),\boldsymbol\nu/\bar\nu,V^{(0)},n_{\mathrm{sh}}\bigr).
\label{supp-eq:initial-gate-inputs}
\end{equation}
Here $\bar\nu$ is the pair mean, $n_{\mathrm{sh}}=7$ (Co) or 9 (Dy), and
$\operatorname{std}$ centers and scales by the population standard deviation.
Norms and denominators are floored at $10^{-8}$. Gating precedes a
near-identity channel map. A width-16 MLP maps $\operatorname{std}(2\mathbf e)$
and $n_{\mathrm{sh}}$ to integer scalars. Increment gates add the invariants
in Sec.~\ref{supp-sec:spinor-factorization}, with pair-shared multipliers at fixed $j$.

Before contraction, angular expansion increases tensor-channel multiplicities,
\begin{equation}
\widetilde h=\sigma_{\mathrm{wd}}\bigl(\iota h+\mathcal W[h\otimes V]\bigr),
\qquad h_{\mathrm b}=\frac{\widetilde h}{\sqrt{1+\|\widetilde h\|_F^2}}.
\label{supp-eq:widening}
\end{equation}
The embedding $\iota$ preserves the original spinors, which alone propagate
between layers. The gate depends on integer scalars and is independent of
the eigenbasis column. Each interaction contracts 4 learned pairs of
$h_{\mathrm b}$, with $j\leq j_{\max}^{\mathrm{init}}+2$ and output rank
$\ell\leq2\ell_{\mathrm{sh}}+1$. Readout uses all allowed products with $V$
and the contraction $h_{\mathrm b}h_{\mathrm b}^\dagger$.

Co outputs use the orthonormal symmetric, traceless Cartesian basis
$\mathsf E_q$ of e3nn~\cite{suppbib-geiger2022e3nn},
$\widehat{\mathbf D}=\sum_{q=-2}^{2}\hat y_q^{(2)}\mathsf E_q$.
For PARASHIFT, let $r_{\ell q}=\|O_q^{(\ell)}\|_F^2/\|O_0^{(\ell)}\|_F^2$,
so $r_{\ell,-q}=r_{\ell q}$ and $r_{\ell0}=1$.
Equal-norm spherical operators use $T_0^{(\ell)}=O_0^{(\ell)}$
and the Condon--Shortley convention~\cite{suppbib-varshalovich1988quantum}.
For $c=U_\ell\hat y^{(\ell)}$, fixing the phase by
$U_\ell Y_{\mathrm r}^{(\ell)}(\hat{\mathbf z})=e_0$, with
$(e_0)_q=\delta_{q0}$, for unit-norm harmonics gives
\begin{equation}
B_0^{(\ell)}=c_0,\qquad
B_q^{(\ell)}=\frac{c_{-q}+(-1)^qc_q}{\sqrt{2r_{\ell q}}},\qquad
B_{-q}^{(\ell)}=\frac{i[(-1)^qc_q-c_{-q}]}{\sqrt{2r_{\ell q}}},\quad q>0.
\label{supp-eq:stevens-output}
\end{equation}
All PARASHIFT neural models use this map and raw Stevens labels, including Stevens factors.

\subsection{Factorized tensor updates}
\label{supp-sec:spinor-factorization}

In exact arithmetic, before learned gates and numerical floors,
Eq.~\eqref{supp-eq:spinor-weights} gives
\begin{equation}
\rho_B=zz^\dagger=\frac{e^{-H_{\mathrm{init}}/T_B}}
{\operatorname{tr}e^{-H_{\mathrm{init}}/T_B}},\qquad
z=\rho_B^{1/2}\Psi.
\label{supp-eq:initializer-gibbs-factor}
\end{equation}
This has the form of a thermal density matrix, with $z$ its factor and $u$
the auxiliary index of a purification~\cite{suppbib-barthel2013thermal}. The analogy
is formal because $H_{\mathrm{init}}$ is approximate and $T_B$ is learned.
The later feature Gram matrix $hh^\dagger$ has no imposed thermal form.

For a Kramers pair, collect angular components and channels into
$h_p\in\mathbb C^{d\times2}$ and define $\rho_p=h_ph_p^\dagger$.
These matrices determine $h_p$ up to $\mathrm U(2)$ and decompose into
integer tensors. Gate inputs follow from
$\|\Pi_jh_p\|_F^2=\operatorname{tr}(\Pi_j\rho_p)$ and
$\|h_p^\dagger h_{p'}\|_F^2=\operatorname{tr}(\rho_p\rho_{p'})$,
where $\Pi_j$ projects onto angular momentum $j$. At fixed geometry and $V$,
write $\delta h_p=\mathcal M h_p$ and let $S_p$ contain the pair-shared gates.
The residual update gives
\begin{equation}
A_p=I+S_p\mathcal M,\qquad h_p^+=A_ph_p,\qquad
\rho_p^+=A_p\rho_pA_p^\dagger.
\label{supp-eq:pair-matrix-recurrence}
\end{equation}
Increment gates use $\mathcal M\rho_p\mathcal M^\dagger$ in these identities,
so $A_p$ is input dependent.
Writing expansion as $\widetilde h=F_Vh$ gives
$\widehat\rho=F_V(\sum_p\rho_p)F_V^\dagger$ and
$\rho_{\mathrm b}=\widehat\rho/(1+\operatorname{tr}\widehat\rho)$.
Learned channel maps give contractions $P_c\rho_{\mathrm b}(P'_c)^\dagger$.
Thus complete pair matrices, including angular and channel cross terms,
reproduce the predictions. At fixed gates, $A=\sum_\alpha a_\alpha T_\alpha$
links tensor-interaction coefficients as $a_\alpha a_\beta^*$.
NequIP+H uses different integer-feature updates
(Sec.~\ref{supp-sec:baseline-electronic-features}).

\subsection{Numerical equivariance and implementation tests}
\label{supp-sec:numerical-checks}

Residuals use $\epsilon(X,Y)=\|X-Y\|_F/\max(\|X\|_F,\|Y\|_F,10^{-30})$.
Each shell uses 3 distorted 6-ligand geometries with synthetic Kramers-paired
eigenvectors, 3 rotations, 3 within-pair gauge changes, and a sign flip.
Networks have 3 layers, 4 channels through rank 6, and 4 contraction pairs,
with nonzero spinor contributions verified.
Maximum float32/complex64 residuals $(\epsilon_h,\epsilon_y)$ are
$(4.96\times10^{-7},1.64\times10^{-6})$ for rotation,
$(3.61\times10^{-7},1.79\times10^{-6})$ for gauge, and $(0,0)$ for sign.
Outputs cover ranks $(2,4)$ for $d$ and $(2,4,6)$ for $f$ shells.
Across 9 float64/complex128 initializer cases, covariance, paired-energy,
and time-reversal residuals are at most $2.82\times10^{-15}$.
Spinor and matrix updates agree within $2.38\times10^{-7}$.
Output-conversion checks at $J=15/2$ and 4 use tolerances $2\times10^{-5}$
(float32) and $10^{-11}$ (float64).

\begin{figure}[!htbp]
\centering
\includegraphics[width=\textwidth]{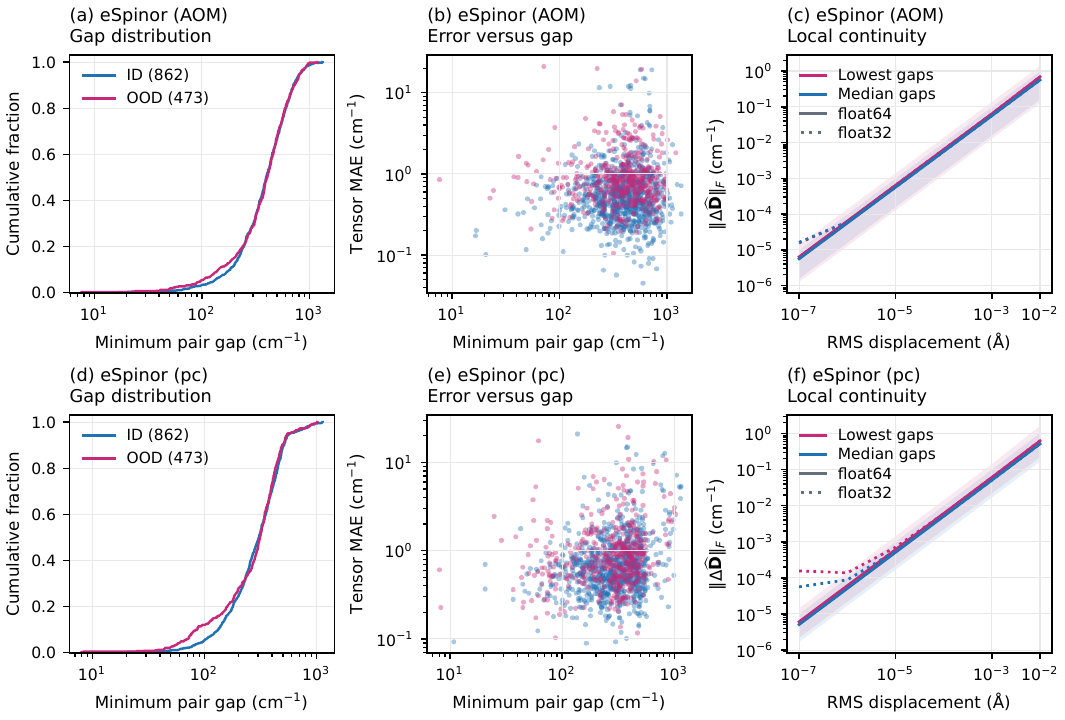}
\caption{Kramers-pair gaps and prediction continuity for the current Co
eSpinor checkpoints with AOM (top) and point-charge (bottom) initialization.
(a), (d) Cumulative gap distributions.
(b), (e) Cartesian tensor MAE per complex.
(c), (f) Median output changes over 24 geometries and 2 directions,
taking the larger change from the 2 displacement signs.
Shading spans the double-precision minimum and maximum.}
\label{supp-fig:kramers-gap}
\end{figure}

We also examine the trained Co initializers on 862 in-distribution and 473
ligand-OOD geometries. For pair energies
$E_p=(\epsilon_{2p-1}+\epsilon_{2p})/2$, the minimum inter-pair gap is
$\Delta_{\min}=\min_p(E_{p+1}-E_p)$.
Its smallest values are 7.62~cm$^{-1}$ for AOM and 8.02~cm$^{-1}$ for
point-charge initialization. Smaller gaps show no systematic increase in
tensor error in Fig.~\ref{supp-fig:kramers-gap}.

For each initializer and split, we perturb the 6 smallest-gap geometries
and 6 near the median in 2 random directions, keeping the metal fixed.
Nonmetal RMS displacements range from $10^{-7}$ to $10^{-2}$~\AA, with
neighbors and ligand masks recomputed. At small steps with unchanged
connectivity, double-precision output changes are proportional to displacement.
Single precision reaches a numerical floor at the smallest steps.
These checks concern local continuity at the sampled finite gaps.

\subsection{Parallel transport for multinuclear systems}
\label{supp-sec:multinuclear-transport}

For same-shell metals, transport between independent eigenbases~\cite{suppbib-cohen2019gauge} uses
\begin{equation}
\Gamma_{b\to a}=\Psi_b^\dagger K_{ba}\Psi_a,\qquad
h_{b\to a}=h_b\Gamma_{b\to a},\qquad
\widetilde h_a=h_a+\sum_{b\in\mathcal N_{\mathrm{metal}}(a)}w_{ab}h_{b\to a}.
\label{supp-eq:multinuclear-connection}
\end{equation}
$K_{ba}$ maps receiver to sender bases ($K_{ba}=I$ in common axes),
and $w_{ab}$ is invariant.
For $K_{ba}'=UK_{ba}U^\dagger$,
$\Gamma_{b\to a}'=Q_b^\dagger\Gamma_{b\to a}Q_a$ and
$h_{b\to a}'=\mathcal D h_{b\to a}Q_a$ give covariance at the receiving site.

\section{Data preprocessing and training}
\label{supp-sec:training-details}

\subsection{Datasets and preprocessing}
\label{supp-sec:cobalt-split}
\label{supp-sec:protocol}

COMPASS filtering retains 14573 of 15559 Set-1 records with complete
ab initio ligand-field entries, including VLFT, and 206 ligand
identifiers~\cite{suppbib-mariano2024compass,suppbib-mariano2024compassdata}.
These entries check completeness only. NumPy's \texttt{default\_rng(42)}
shuffles sorted identifiers, reserving the first 41. Among 4-coordinate
complexes, 8616 use only retained ligands and 473 use only reserved ligands
(40 identities), forming the OOD set. Mixed complexes are excluded.
Shuffling the 8616 structures from filtered dataset order with
\texttt{default\_rng(43)} gives 862 validation, 862 in-distribution test,
and 6892 training structures, in that order.

Dy25k filtering retains 22,726 complexes with 7 non-H neighbors within
3~\AA\ of Dy, comprising 5 water O atoms and 1 donor from each of
2 organic ligands, both O or both N~\cite{suppbib-frangoulis2026dysmm,suppbib-frangoulis2026dy25kdata}.
Structures sharing a source identifier or ligand heavy-atom connectivity
form the same ligand family, including protonation variants. An 80/10/10
split of these families, stratified by donor element, gives 18,231 available
training, 2244 validation, and 2251 test complexes. Training subsets of
1k, 2k, 4k, and 8k are nested. The test set contains 1560 O-donor and
691 N-donor complexes.

PARASHIFT provides Dy CASSCF-SO labels along the trajectory of the deuterated
Gd analogue in D$_2$O~\cite{suppbib-thompson2026parashift,suppbib-chilton2025parashiftdata}.
At 1~fs spacing, the zero-based frame intervals are
\begin{equation}
\mathcal I_{\mathrm{train}}=[0,N),\quad N\in\{500,1000,2000,4000,8000\},\qquad
\mathcal I_{\mathrm{val}}=[50000,51000),\qquad
\mathcal I_{\mathrm{test}}=[51000,90000).
\label{supp-eq:dy-splits}
\end{equation}
Preprocessing selects periodic images nearest the metal and retains atoms
within 6~\AA, including solvent. Coordinates and targets share laboratory axes.
PARASHIFT neural fits use seeds 0, 1, and 42 on the fixed split.
Errors are averaged over fits. Bands show sample standard deviations,
faint points show individual fits, and larger markers show means.

\subsection{Neural models and optimization}
\label{supp-sec:baseline-electronic-features}

Table~\ref{supp-tab:training-settings} gives the settings for NequIP,
NequIP+H, and eSpinor. Graphs include hydrogens.

\begin{table}[!htbp]
\centering
\caption{NequIP, NequIP+H, and eSpinor settings for the 3 benchmarks,
fixed across training budgets.}
\label{supp-tab:training-settings}
\begin{tabular*}{\textwidth}{@{\extracolsep{\fill}}lccc@{}}
\toprule
Setting & COMPASS & Dy25k & PARASHIFT\\
\midrule
Layers / maximum integer rank / channels per rank & \multicolumn{3}{c}{3 / 6 / 4}\\
Radial basis size / cutoff order & \multicolumn{3}{c}{8 / 6}\\
NequIP+H initializer & AOM & Point charge & Point charge\\
eSpinor initializer & AOM or point charge & Point charge & Point charge\\
Graph cutoff (\AA) / neighbor normalization & 6 / 10 & 4 / 6 & 4 / 6\\
Integer radial MLP depth / width & 3 / 64 & 1 / 128 & 1 / 128\\
eSpinor radial hidden width & 64 & 4 & 4\\
Batch size & 32 & 64 & 64\\
Initial $T_B$ (cm$^{-1}$), NequIP+H & 200 & 7500 & 200\\
Initial $T_B$ (cm$^{-1}$), eSpinor & 200 & 7500 & 7500\\
Peak / final learning rate & \multicolumn{3}{c}{$2\times10^{-3}$ / $10^{-6}$}\\
Warmup / maximum epochs / stopping patience & \multicolumn{3}{c}{10 / 300 / 100}\\
Weight decay / gradient norm limit / EMA decay & \multicolumn{3}{c}{$10^{-4}$ / 1 / 0.99}\\
\bottomrule
\end{tabular*}
\end{table}

NequIP+H and eSpinor fit their initializers separately, using reference
magnetic parameters only as targets. At each layer, NequIP+H gates the
weighted initial eigenvectors $z$ using current integer scalars and
Eq.~\eqref{supp-eq:initial-gate-inputs}. The invariant diagonal gate $S_{\mathrm H}$
has equal values within each Kramers pair. Its gauge-invariant projector
sum and normalized channel contractions are
\begin{equation}
\rho_{\mathrm H}=(zS_{\mathrm H})(zS_{\mathrm H})^\dagger,\qquad
b_{\mathrm H}=\frac{zS_{\mathrm H}}{\sqrt{1+\|zS_{\mathrm H}\|_F^2}},\qquad
C_c=(P_cb_{\mathrm H})(P'_cb_{\mathrm H})^\dagger.
\label{supp-eq:baseline-contraction-details}
\end{equation}
Maps $P_c,P'_c$ multiply each $j$ subspace by a learned complex scalar.
Projection gives real/imaginary tensor channels without expansion.
Each layer recomputes these tensors from $z$ and couples them to $V$,
which propagates the electronic information. PARASHIFT readout multiplicities at ranks $(2,4,6)$ are
$(3,3,3)$ for NequIP+H and $(43,64,73)$ for eSpinor.
The point-charge eSpinor initializer precedes all 3 convolutions, while Co
AOM follows the first convolution and leaves 2 spinor updates.

For Dy25k, a Dy-site readout with hidden widths $(64,64)$, SiLU activations,
and layer normalization maps scalar features to the 7 excitation energies.
NequIP+H and eSpinor also supply electronic contraction invariants.
The readout learns residuals about the per-level training means.

All 3 models minimize the normalized prediction error
\begin{equation}
\mathcal L=\frac1{|\mathcal T|}\sum_{r\in\mathcal T}
\frac{\langle\|\hat{\mathbf y}_r-\mathbf y_r\|_2^2\rangle_{\mathcal B}}
{d_r s_r^2}.
\label{supp-eq:training-loss}
\end{equation}
Here $r\in\mathcal T$ labels prediction heads, $\mathbf y_r$ and
$\hat{\mathbf y}_r$ are reference and predicted component vectors, and
$\langle\cdot\rangle_{\mathcal B}$ is the batch mean.
For tensor targets, $d_r=9$ for Co and $5$, $9$, and $13$ for PARASHIFT ranks
$2$, $4$, and $6$, with $s_r$ the training-target standard deviation over
samples and components. For Dy25k, $d_r=7$ and
$s_r^2=\frac17\sum_{k=1}^7\operatorname{Var}_{\mathrm{train}}(\Delta E_{0k})$.
All scales are floored at $10^{-6}$ in target units.
Both Dy benchmarks use fixed Bessel functions and a polynomial cutoff~\cite{suppbib-gasteiger2020dimenet}.
Linear warmup starts at 0.01 times the peak rate, followed by cosine decay.
AdamW~\cite{suppbib-loshchilov2019adamw} excludes biases and 1-dimensional parameters
from weight decay. The lowest validation loss selects the checkpoint,
evaluated using exponential-moving-average weights.

Features use float32/complex64 in PyTorch~\cite{suppbib-paszke2019pytorch},
with higher-precision angular tables, e3nn~\cite{suppbib-geiger2022e3nn}, and
cuEquivariance tensor products~\cite{suppbib-nvidia2026cuequivariance}.
Co training used a 32~GB V100 on Taiwania 2 (TWCC), with PyTorch
2.9.1+cu126 for eSpinor and NequIP+H, 2.8.0+cu128 for NequIP, and
cuEquivariance 0.8.0 for eSpinor. PARASHIFT inference used an H200 with
naive tensor-product kernels.

\subsection{Descriptor baselines}
\label{supp-sec:descriptor-baselines}

eqSNAP uses MolForge bispectra through $2j_{\max}=8$ and a 4~\AA\
cutoff~\cite{suppbib-nguyen2022eqsnap,suppbib-briganti2025spinphonon}. For PARASHIFT, species-pooled
features and rankwise ridge fits are
\begin{equation}
X_{\ell m,si}=\sum_{a:s(a)=s}b_i(a)\sum_{b\ne a}
f_{\mathrm{cut}}(r_{ab})Y_{\ell m}(\hat{\mathbf r}_{ab}),\qquad
\widehat T_{\ell m}=\sum_{s,i}\beta_{si}^{(\ell)}X_{\ell m,si},\qquad
\mathcal L=\|X\beta-y\|_2^2+\lambda^2\|\beta\|_2^2.
\label{supp-eq:ridge-objective}
\end{equation}
Weights are shared over $m$. With 5 species and 55 bispectra, ranks $(2,4,6)$
use 825 real weights, stored as 1650 real/imaginary entries.
PARASHIFT fits raw Stevens units with $\lambda=0.1$ at every budget.
Co uses metal-centered descriptors and validation tensor RMSE to select
$\lambda=10$ from $\{0.01,0.1,1,10,100\}$.
Dy25k uses rank-0 descriptors and training-data normalization of features
and targets. Validation selects all-atom pooling and $\alpha=10^{-3}$ in
$N_{\mathrm{train}}^{-1}\|X\beta-y\|_2^2+\alpha\|\beta\|_2^2$.

GM-NN~\cite{suppbib-zaverkin2022gmnn} uses 5 radial functions with 7 Gaussians
each, hidden widths $(512,512)$, and linear learning-rate decay over at most
1000 epochs. Co uses an 8~\AA\ cutoff, batch size 32, initial rate 0.02,
and a 5-component tensor loss, with checkpoint selection by validation
tensor MAE. Dy25k uses a 4~\AA\ cutoff and batch size 64. Its 7 atomic
outputs are summed and divided by $\sqrt{\bar N}$ before applying learned
per-level scales and offsets, where $\bar N$ is the mean training atom count.
Adam minimizes
Eq.~\eqref{supp-eq:training-loss}, starting at 0.02 for network parameters
and 0.025 for output scales and offsets. Validation loss selects the
checkpoint, with patience 100.

\section{Cobalt tensor errors and anisotropy screening}
\label{supp-sec:cobalt-details}
\label{supp-sec:cobalt-screening}

We remove the trace from the released matrices,
$\mathbf D=\mathbf D_{\mathrm{raw}}-\operatorname{tr}(\mathbf D_{\mathrm{raw}})I_3/3$.
This preserves level splittings. In principal axes, $z$ follows the largest
absolute eigenvalue, $D=3D_{zz}/2$, and $E=(D_{xx}-D_{yy})/2$~\cite{suppbib-boca2004zfs}.
We evaluate $|E|$ because exchanging $x,y$ reverses its sign.
Per-complex tensor errors are
$e_i=\frac19\sum_{\alpha,\beta=x,y,z}|\widehat D_{i,\alpha\beta}-D_{i,\alpha\beta}|$.
Table~\ref{tab:cobalt-tensors} averages $e_i$, $|\widehat D_i-D_i|$,
and $\bigl||\widehat E_i|-|E_i|\bigr|$ over complexes.

Spearman correlation compares predicted and reference ranks of $D$.
Shortlists $\mathcal S_k,\widehat{\mathcal S}_k$ contain their $k$ most
negative values, with recovery
$r_k=|\widehat{\mathcal S}_k\cap\mathcal S_k|/k$ for
$k=\lceil fN\rceil$ and $f\in\{.01,.02,.05,.10,.20,.50\}$.
Figure~\ref{supp-fig:cobalt-screening} compares candidate rankings by axial
anisotropy. Tensor errors avoid the axis-reassignment
sensitivity of scalar $D$ near the rhombic limit.

\par\medskip\noindent
\begin{minipage}{\textwidth}
\centering
% Trim only the empty vertical margins, preserving the full plot width.
\includegraphics[width=\textwidth,trim=0 8bp 0 4bp,clip]{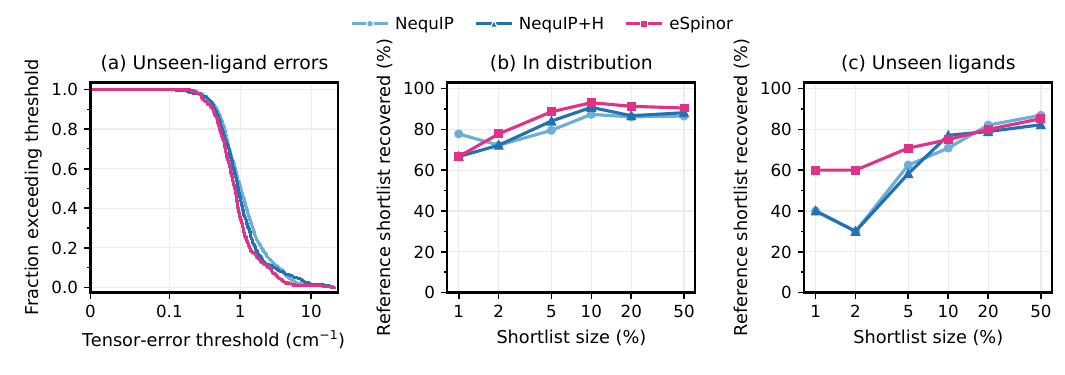}
\captionof{figure}{Cobalt screening for the neural models. eSpinor uses AOM.
(a) OOD fraction exceeding tensor-error threshold $e_i$, on an axis linear
up to 0.1~cm$^{-1}$ and logarithmic above.
(b), (c) Shortlist recovery for 862 in-distribution and 473 OOD complexes.}
\label{supp-fig:cobalt-screening}
\end{minipage}
\par\medskip

\section{Dysprosium excitation spectra and ligand screening}
\label{supp-sec:dy-spectra-details}

Figure~\ref{supp-fig:dy-spectra-learning} shows learning curves and shortlist
recovery as defined in Sec.~\ref{supp-sec:cobalt-screening}. At 8k, eSpinor has
the lowest spectrum MAE among all 5 models for both O and N donors
(15.42 and 22.99~cm$^{-1}$). For the top 23 complexes by total splitting,
NequIP+H recovers 21, compared with 18 for eSpinor and 16 for NequIP.

\par\medskip\noindent
\begin{minipage}{\textwidth}
\centering
% Trim only the empty vertical margins, preserving the full plot width.
\includegraphics[width=\textwidth,trim=0 8bp 0 2bp,clip]{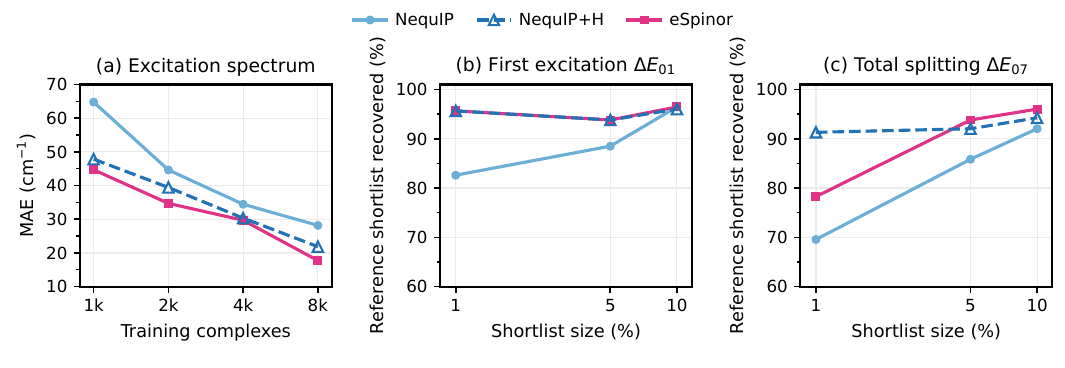}
\captionof{figure}{Dy25k ligand-OOD prediction on 2251 test complexes.
(a) Spectrum MAE versus training-set size.
(b), (c) Recovery of the reference top 1\%, 5\%, and 10\% by first
excitation and total splitting for the 8k models.}
\label{supp-fig:dy-spectra-learning}
\end{minipage}
\par\medskip

For the combined thresholds in Table~\ref{tab:dy-spectra}, the
eSpinor false positive lies just below the total-splitting cutoff, with
reference $\Delta E_{07}=1485.72$~cm$^{-1}$.
Predictions are evaluated without sorting or clipping. Negative or
out-of-order excitation energies occur in some predicted spectra, but
not among the candidates selected by the 8k models.

\section{Dysprosium dynamics and additional results}
\label{supp-sec:dysprosium-details}

\subsection{Propagation and coordinate conventions}
\label{supp-sec:spin-dynamics-protocol}
\label{supp-sec:computational-cost}

Following \apxref{app:spin-relaxation} and
PARASHIFT~\cite{suppbib-thompson2026parashift}, water propagation uses 78 nonoverlapping
500-frame test windows, $\Delta t=1$~fs, 100 equally
weighted Fibonacci field orientations, $T=300$~K, and $\tau_R=139$~ps.
Fields are $\mathcal B=\{1,1.5,2,3,10,20\}$~T, with low-field averages over
$\mathcal B_{\mathrm{MRI}}=\{1,1.5,2,3\}$~T. Each model uses
\begin{equation}
\rho_0=\frac{e^{-H_0/(\bar k_{\mathrm B}T)}}{\operatorname{Tr}e^{-H_0/(\bar k_{\mathrm B}T)}},
\qquad U_n=e^{-i2\pi cH_n\Delta t},\qquad P_0=I,\qquad P_{n+1}=U_nP_n.
\label{supp-eq:thermal-state}
\end{equation}
Here $H_n=H_{\mathrm{eff}}(t_0+n\Delta t)$, $\bar k_{\mathrm B}=0.6950348$
cm$^{-1}$\,K$^{-1}$, and time in the exponential is in seconds.
Hermitian diagonalization uses float64/complex128, subtracting the lowest
energy in the thermal exponential. Electron $g_{zz}$ uses laboratory axes
and a fixed field, with 499 steps and 500 lags.

Nuclear correlations and susceptibility use minimum-image, metal-centered
coordinates. Orient donor-plane unit normals $\boldsymbol\nu_A$ from the
axial-N centroid toward the O centroid and define the molecular frame by
\begin{equation}
\mathbf e_z=\widehat{\boldsymbol\nu_{\mathrm O}+\boldsymbol\nu_{\mathrm{N,ax}}+
\boldsymbol\nu_{\mathrm{N,eq}}},\quad
\mathbf e_x=\widehat{\mathbf r_{\mathrm{N,eq},1}-(\mathbf r_{\mathrm{N,eq},1}\cdot\mathbf e_z)\mathbf e_z},
\quad \mathbf e_y=\mathbf e_z\times\mathbf e_x,\quad R=(\mathbf e_x,\mathbf e_y,\mathbf e_z)^T.
\end{equation}
Hats normalize vectors. Water donor indices are O $(1,3,5)$, axial N
$(721,722,723)$, and equatorial N $(724,725,726)$, counted from 1.
The same $R$ rotates coefficients and Dy--proton vectors, with field
$\hat{\mathbf n}(t)=R(t)R(t_0)^T\hat{\mathbf n}(t_0)$.

Exporting 39k predictions averaged 709.6~s for 8k eSpinor models on 1 H200
(batch size 8, naive tensor products), including setup and I/O but excluding
training and propagation. This corresponds to 18.2~ms per geometry averaged
over the export. For context, 2 approximate OpenMolcas CASSCF-SO
jobs~\cite{suppbib-galvan2019openmolcas} took 8411 and 5137~s per geometry, including
input preparation and parameter extraction, on 72 NVIDIA Grace CPU cores
on DeltaAI~\cite{suppbib-bode2025deltaai}. Each used 1 MPI process and 72 OpenMP
threads, with no GPU computation. The setup used CAS(9,7), 18 equally
weighted sextet roots, RASSI SOC, and ANO-RCC bases with VTZP on Dy,
VDZP on O/N, and VDZ elsewhere. Compared with the
benchmark reference protocol, it used placeholder solvent charges
($-0.68$ on O and $+0.34$ on H) and VDZP on all O/N atoms instead of only
donors. These jobs provide a cost scale for electronic-structure evaluation,
not timings for generating the benchmark labels or a controlled speedup ratio.

\subsection{Prediction errors and propagation diagnostics}
\label{supp-sec:coefficient-results}
\label{supp-sec:correlations}
\label{supp-sec:curve-errors}
\label{supp-sec:blocks}
\label{supp-sec:hybrid-propagation}

\begin{figure}[!htbp]
\centering
\includegraphics[width=\textwidth]{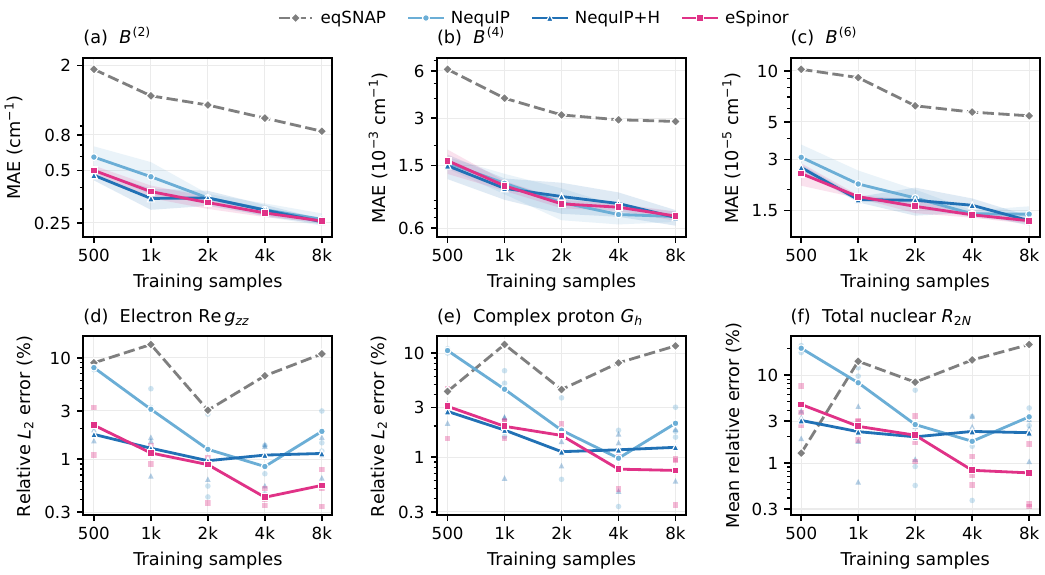}
\caption{Water learning curves on 39k test frames.
(a)--(c) Crystal-field MAEs.
(d), (e) Real electron and complex proton correlations over all 6 fields.
(f) Total $R_{2N}$ errors over $1$--$3$~T. Proton quantities average H3/H4/H5.
$R_{1N}$ gives the same model ordering at every budget.}
\label{supp-fig:parashift-lc}
\label{supp-fig:observable-learning}
\end{figure}

For $N_{\mathrm{test}}$ geometries and the 500 explicit correlation samples
$\mathbf C$, without fitted tails, we use
\begin{equation}
\mathrm{MAE}_\ell=\frac{\sum_{i,q}|\widehat B_{q,i}^{(\ell)}-B_{q,i}^{(\ell)}|}
{N_{\mathrm{test}}(2\ell+1)},\qquad
\epsilon_C^{\mathcal P}=100\frac{\|\mathcal P\mathbf C_{\mathrm{model}}-
\mathcal P\mathbf C_{\mathrm{ref}}\|_2}{\|\mathcal P\mathbf C_{\mathrm{ref}}\|_2}.
\label{supp-eq:curve-error}
\end{equation}
Here $\mathcal P\in\{\operatorname{Re},\operatorname{Im},\mathrm{id}\}$ selects
real, imaginary, or complex correlations. Shape errors normalize each curve
by its own $\operatorname{Re}C(0)$. Errors average 6 fields equally and,
for $G_h$, H3/H4/H5. Figure~\ref{supp-fig:observable-learning} includes all budgets
and models. The normalized real-correlation advantage persists at 1k--8k
and in each 26-window segment starting at 51k, 64k, and 77k at 8k.
At 8k, imaginary-correlation errors favor NequIP+H over eSpinor
($1.264\%$ versus $1.534\%$).

To isolate propagation from initial-state error, we combine the reference
thermal state with each model's propagator,
\begin{equation}
c_{RM}(\tau)=\operatorname{Tr}[\rho_{0,R}J_\parallel P_M^\dagger J_\parallel P_M],
\qquad e_{wo}(\tau)=\operatorname{Re}[c_{RM}^{w,o}(\tau)-c_{RR}^{w,o}(\tau)].
\label{supp-eq:hybrid-error-decomposition}
\end{equation}

\begin{table}[!htbp]
\centering
\caption{Dimensionless propagation RMS errors at 8k with the reference
initial state. Columns follow Eq.~\eqref{supp-eq:propagation-error-averaging}.}
\label{supp-tab:dy-propagation-diagnostics}
\begin{tabular*}{\textwidth}{@{\extracolsep{\fill}}lrrrrrr@{}}
\toprule
 & \multicolumn{3}{c}{3 T} & \multicolumn{3}{c}{20 T}\\
\cmidrule(lr){2-4}\cmidrule(l){5-7}
Model & $E_{\mathrm{individual}}$ & $E_{\mathrm{orientations}}$ & $E_{\mathrm{ensemble}}$
 & $E_{\mathrm{individual}}$ & $E_{\mathrm{orientations}}$ & $E_{\mathrm{ensemble}}$\\
\midrule
NequIP & 0.7270 & 0.4184 & 0.1962 & 0.7009 & 0.3104 & 0.1859\\
NequIP+H & 0.6929 & 0.3706 & 0.1129 & 0.6718 & 0.2701 & 0.1143\\
eSpinor & 0.7195 & 0.3795 & 0.0559 & 0.6882 & 0.2419 & 0.0572\\
\bottomrule
\end{tabular*}
\end{table}

Here $R,M$ denote reference and model, with identical Zeeman terms.
For windows $w$, orientations $o$, and arithmetic means,
\begin{equation}
E_{\mathrm{individual}}=\langle e^2\rangle_{w,o,\tau}^{1/2},\qquad
E_{\mathrm{orientations}}=\langle\langle e\rangle_o^2\rangle_{w,\tau}^{1/2},\qquad
E_{\mathrm{ensemble}}=\langle\langle e\rangle_{w,o}^2\rangle_\tau^{1/2}.
\label{supp-eq:propagation-error-averaging}
\end{equation}
Tests use all windows and orientations at 8k and 3/20~T. Unmixed runs
reproduce saved correlations within $7\times10^{-14}$.
Table~\ref{supp-tab:dy-propagation-diagnostics} shows comparable individual errors
but a smaller ensemble-mean signed error for eSpinor than for NequIP+H.
Here the mean residual $b(\tau)=\langle e(\tau)\rangle_{w,o}$ measures the
systematic error in the sampled ensemble, and
$E_{\mathrm{ensemble}}=\langle b(\tau)^2\rangle_\tau^{1/2}$.

\subsection{Relaxation calculations and sensitivity}
\label{supp-sec:rates}
\label{supp-sec:proton-correlations}
\label{supp-sec:fits}
\label{supp-sec:rate-construction}
\label{supp-sec:rate-errors}
\label{supp-sec:sensitivity}

Each proton's real $G_p$ is fitted separately with the PARASHIFT
form~\cite{suppbib-thompson2026parashift},
\begin{equation}
C_{\mathrm{fit}}(\tau)=\begin{cases}
C_{\mathrm G}(\tau)=C(0)e^{-(\tau/T_{\mathrm G})^2},&\tau<\tau_0,\\
C_{\mathrm S}(\tau)=C(\tau_0)e^{-((\tau-\tau_0)/T_{\mathrm S})^\beta},&\tau\geq\tau_0.
\end{cases}
\label{supp-eq:fit}
\end{equation}
For each $\tau_0=k\Delta t$, $k=2,\ldots,L-3$, we anchor the late branch
at the observed $C(\tau_k)>0$. SciPy~\cite{suppbib-virtanen2020scipy} minimizes
$[\sum_{n=0}^{k}(C_{\mathrm G}(\tau_n)-C(\tau_n))^2+
\sum_{n=k}^{L-1}(C_{\mathrm S}(\tau_n)-C(\tau_n))^2]/[LC(0)^2]$,
with $\Delta t/100\leq T_{\mathrm G},T_{\mathrm S}\leq1000\tau_{L-1}$ and
$0.1\leq\beta\leq2$. We choose the converged fit with the smallest objective.
Fitting $\operatorname{Re}g_{zz}$ characterizes electron decay only.
At 8k, eSpinor leads $T_{\mathrm G},T_{\mathrm S}$ accuracy and NequIP leads
$\beta,\tau_0$.

The explicit complex $G_p$ is retained through 499~fs and followed by the real
fitted tail from 500~fs. We include the first tail sample below
$\eta\operatorname{Re}G_p(0)$, with $\eta=10^{-3}$, then set the tail to zero.
This cutoff follows our reconstruction of the archive.
The spectral integral is evaluated with equal endpoint weights,
$\mathcal J_p(\omega)=\Delta t\sum_{n=0}^{n_c}\widetilde G_p(\tau_n)
e^{-\tau_n/\tau_R-i\omega\tau_n}$, with time in seconds.
This integral gives the dipolar rates derived in \apxref{app:spin-relaxation}.
Proton-group correlations $G_h=\frac13\sum_{p\in h}G_p$ are used only for
curve comparisons, so averaging does not precede the proton-resolved fits.

For the mean shielding $\overline{\boldsymbol\sigma}_p$ defined in that appendix,
let $S_p$ and $A_p$ be its symmetric and antisymmetric parts. The invariants
$a_p^2=2\|A_p\|_F^2$ and
$b_p^2=\frac32\|S_p-\operatorname{tr}(S_p)I/3\|_F^2$ give
\begin{equation}
\begin{aligned}
J_1(\omega)&=\frac{\tau_R}{1+9\omega^2\tau_R^2},&
J_2(\omega)&=\frac{\tau_R}{1+\omega^2\tau_R^2},\\
R_{1N,p}^{\mathrm{Curie}}&=\omega_N^2\left[\frac{a_p^2}{2}J_1(\omega_N)
+\frac{2b_p^2}{15}J_2(\omega_N)\right],&
R_{2N,p}^{\mathrm{Curie}}&=\omega_N^2\left[\frac{a_p^2}{4}J_1(\omega_N)
+\frac{b_p^2}{45}(4\tau_R+3J_2(\omega_N))\right].
\end{aligned}
\end{equation}
No empirical scale factor is applied. Reference Curie rates use archived
CASSCF-SO susceptibilities. Model susceptibilities use the zero-field
Van Vleck expression within $J=15/2$~\cite{suppbib-vanvleck1932susceptibilities}.
For energies $\mathcal E_n$, thermal weights $p_n$, $\beta_T=(\bar k_{\mathrm B}T)^{-1}$,
and $M_{\alpha,nm}=\langle n|g_JJ_\alpha|m\rangle$,
\begin{equation}
\begin{aligned}
F^\chi_{nm}&=\begin{cases}(p_n-p_m)/(\mathcal E_m-\mathcal E_n),
&|\mathcal E_m-\mathcal E_n|\geq\delta_\chi,\\
\beta_Tp_n,&|\mathcal E_m-\mathcal E_n|<\delta_\chi,\end{cases}&
\chi_{\mathrm{mol},\alpha\beta}&=\Lambda_\chi\sum_{n,m}F^\chi_{nm}
\operatorname{Re}[M_{\alpha,nm}M_{\beta,nm}^*],\\
\delta_\chi&=10^{-6}~\mathrm{cm}^{-1},&
\boldsymbol\chi_{\mathrm{SI}}&=\frac{4\pi\,10^{-6}}{N_A}\boldsymbol\chi_{\mathrm{mol}}.
\end{aligned}
\end{equation}
Here $\Lambda_\chi=0.26074098196$~cm$^3$\,cm$^{-1}$\,mol$^{-1}$ gives cgs
molar susceptibility, and $N_A$ is Avogadro's constant.
Errors against the archive include the ground-multiplet approximation.

Total rates are averaged within each proton environment before taking errors.
For $a=1,2$ and $\mathcal H=\{\mathrm{H3,H4,H5}\}$,
\begin{equation}
\overline R_{a,h}=\tfrac13\sum_{p\in h}R_{aN,p},\quad
\epsilon_{R_a}(B)=\frac{100}{3}\sum_{h\in\mathcal H}
\frac{|\overline R_{a,h}^{\mathrm{model}}(B)-\overline R_{a,h}^{\mathrm{ref}}(B)|}
{|\overline R_{a,h}^{\mathrm{ref}}(B)|},\quad
\epsilon_{R_a}^{\mathrm{MRI}}=\tfrac14\sum_{B\in\mathcal B_{\mathrm{MRI}}}\epsilon_{R_a}(B).
\label{supp-eq:rate-error}
\end{equation}
Component errors use their own reference mechanism in the denominator.

Susceptibility errors compare the trajectory and its mean,
\begin{equation}
\epsilon_\chi^{\mathrm{traj}}=100\sqrt{
\frac{\sum_i\|\widehat\chi_i-\chi_i\|_F^2}{\sum_i\|\chi_i\|_F^2}},\qquad
\epsilon_\chi^{\mathrm{mean}}=100\frac{\|\overline{\widehat\chi}-\overline\chi\|_F}{\|\overline\chi\|_F}.
\label{supp-eq:chi-errors}
\end{equation}
\begin{figure}[!t]
\centering
\includegraphics[width=\textwidth]{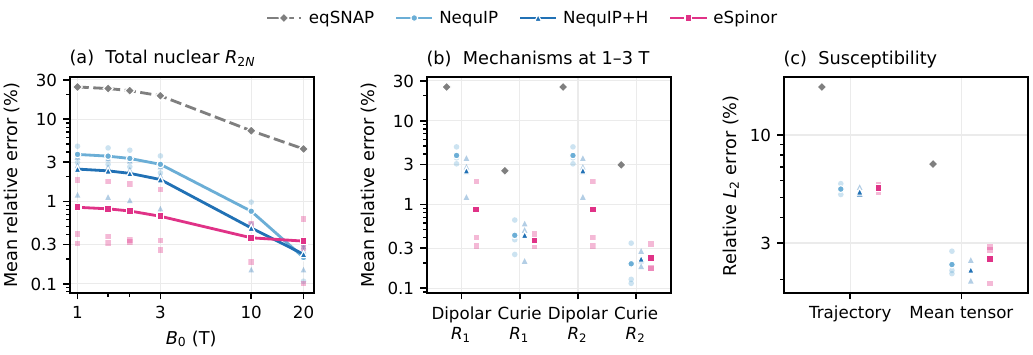}
\caption{Field and mechanism comparisons at 8k.
(a) Total $R_{2N}$ errors over H3/H4/H5 at each field.
(b) Dipolar and Curie errors over $1$--$3$~T, normalized by their own reference
mechanism. (c) Susceptibility errors from Eq.~\eqref{supp-eq:chi-errors}.}
\label{supp-fig:field-mechanisms}
\end{figure}

Figure~\ref{supp-fig:field-mechanisms} compares field-dependent rates, their
mechanisms, and susceptibility.

To check cancellation, split each rate into explicit-integral, fitted-tail,
and Curie contributions, $R_a=\sum_X R_a^X$.
The signed errors $\delta_a^X=100(\widehat R_a^X-R_{a,\mathrm{ref}}^X)/R_{a,\mathrm{ref}}$
then add exactly. For eqSNAP, explicit-integral and tail errors compensate
at 500 samples but reinforce at 8k. With fits fixed, varying $\eta$ from
$10^{-3}$ to $10^{-5}$ and using matched $J=15/2$ reference susceptibility
gives 4 combinations. Mean rate errors change by less than 0.10 percentage
points for every model, with unchanged 8k low-field ranking. Longer
trajectories are needed to validate the extrapolated tails.

\subsection{Transfer from water to methanol}
\label{supp-sec:meoh-transfer}

Water-trained models transfer without retraining or selection on MeOH.
Propagation uses the first 82,000 of 82,300 matched frames in 164 windows.
Susceptibility uses all frames, with archived values for reference Curie rates.
The water protocol is unchanged except $\tau_R=75$~ps and one-based donor
indices O $(1,3,5)$, axial N $(697,698,699)$, and equatorial N $(700,701,702)$.

\par\medskip\noindent
\begin{minipage}{\textwidth}
\centering
\includegraphics[width=\textwidth]{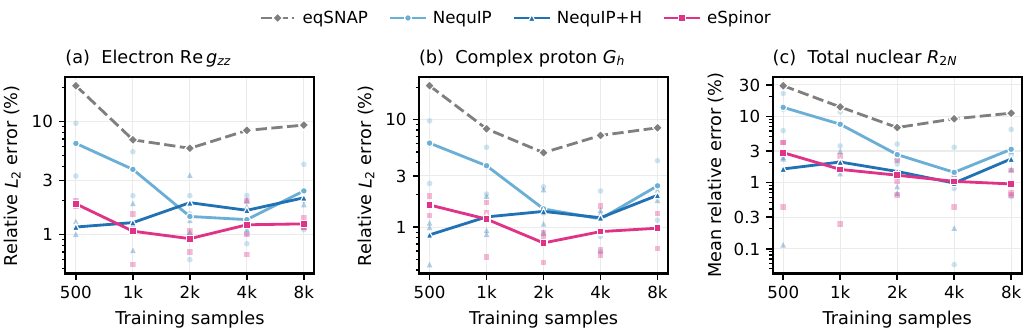}
\captionof{figure}{Water-to-MeOH transfer.
(a), (b) Real electron-spin and complex proton-correlation errors over all
6 fields. (c) Total $R_{2N}$ errors averaged over H3/H4/H5 and $1$--$3$~T.
$R_{1N}$ gives the same model ordering at every budget.
Table~\ref{tab:parashift-transfer} gives the 8k comparison.}
\label{supp-fig:meoh-transfer}
\end{minipage}
\par\medskip

\twocolumngrid
\renewcommand{\bibsection}{\section*{References}}
% REVTeX labels the final bibliography item. Keep this label separate
% from the main bibliography when both reference lists share one document.
\begingroup
\let\arxivbibliographylabel\label
\renewcommand{\label}[1]{\arxivbibliographylabel{supp-bib-#1}}

\endgroup

\end{document}